\documentclass{aa}
\usepackage{epstopdf}
\usepackage{txfonts}
\usepackage{graphicx}
\usepackage{natbib}

\bibpunct{(}{)}{;}{a}{}{,}

\usepackage{color}
\newcommand{\kms}{\,{\rm km\,s}^{-1}}

\newcommand{\Lsun}{L_\odot}
\newcommand{\Msun}{M_\odot}

\newcommand{\as}{\ifmmode {^{\scriptscriptstyle\prime\prime}}
        \else $^{\scriptscriptstyle\prime\prime}$\fi}

\begin{document}
\title{The masses of young stars: \\
CN as a probe of dynamical masses.\thanks{Based on observations carried out with the IRAM Plateau de Bure
interferometer. IRAM is supported by INSU/CNRS (France), MPG (Germany) and IGN (Spain).}
}

\author{S.Guilloteau \inst{1,2}, M.Simon \inst{3}, V.Pi\'etu \inst{4}, E.Di Folco \inst{1,2},
A.Dutrey\inst{1,2}, L.Prato \inst{5}, E.Chapillon \inst{6,1,2,4} }
%
%
\institute{
Univ. Bordeaux, LAB, UMR 5804, F-33270, Floirac, France
\and
CNRS, LAB, UMR 5804, F-33270 Floirac, France\\
  \email{[name]@obs.u-bordeaux1.fr}
\and Department of Physics and Astronomy, Stony Brook University, Stony Brook, NY 11794-3800, USA
\and IRAM, 300 rue de la piscine, F-38406 Saint Martin d'H\`eres, France
  \email{[name]@iram.fr}
\and  Lowell Observatory, 1400 West Mars Hill Road, Flagstaff, AZ 86001, USA
 \email{lprato@lowell.edu}
\and Academia Sinica Institute of Astronomy and Astrophysics, P.O. Box 23-141, Taipei 10617, Taiwan 
}

\offprints{S.Guilloteau, \email{Stephane.Guilloteau@obs.u-bordeaux1.fr}}

\date{Received / Accepted } %
\authorrunning{Guilloteau et al.} %
\titlerunning{CN as a mass tracer of PMS stars.}

\abstract
{}
{We attempt to determine the masses of single or multiple young T Tauri and HAeBe stars from
the rotation of their Keplerian disks.}
{We used the IRAM PdBI interferometer to perform arcsecond resolution images
of the CN N=2-1 transition with good spectral resolution. Integrated
spectra from the 30-m radiotelescope show that CN is relatively unaffected
by contamination from the molecular clouds.  Our sample includes 12 sources, among
which isolated stars like DM Tau and MWC 480 are used to demonstrate the
method and its accuracy. We derive the dynamical mass by fitting a
disk model to the emission, a process giving M/D the mass to distance ratio.
We also compare the CN results with higher
resolution CO data, that are however affected by contamination.
}
{All disks are found in nearly perfect Keplerian rotation. We determine accurate masses for 11 stars, 
in the mass range 0.5 to 1.9$\Msun$.
The remaining one, DG Tau B, is a
deeply embedded object for which CN emission partially arises from the outflow.
With previous determination, this leads to 14 (single) stars with dynamical masses. Comparison
with evolutionary tracks, in a distance independent modified HR diagram,
show good overall agreement (with one exception, CW Tau), and
indicate that measurement of effective temperatures are the limiting factor.
The lack of low mass stars in the sample does not allow to distinguish between
alternate tracks.
}
{}


\maketitle{}

\section{Introduction}

To understand the diversity among the many known planetary systems
it is important to study the evolution of their proto-planetary disks
and to establish a reliable clock for the very early ($< 10$\,Myr) phases.
Ages of the young stars that host the disks can provide the
clocks.
Unfortunately, the age of individual stars is not directly observable, and must rely 
on the comparison between the observed stellar properties and theoretical models of 
early stellar evolution, a model dependent derivation. From an observational point 
of view, stars can be characterized by their mass $M_*$, radius $R_*$, luminosity 
$L_*$ (or the equivalent combinations including effective temperature $T_\mathrm{eff}$ 
and surface gravity $g$), and detailed spectrum. Ages of young stars are usually derived 
through their location in, for example, an HR ($L_* \mathrm{~vs~} T_\mathrm{eff}$) diagram.

The existing stellar evolution models \citep{Baraffe+etal_1998,Dantona+etal_1994,Dantona+Mazzitelli_1997,
Palla+Stahler_1999,Siess+etal_2000},  the Y2 models from \citet{Demarque+etal_2004,Demarque+etal_2008},
the Dartmouth models \citep{Dotter+etal_2008}, and the Pisa tracks \citep{Tognelli+etal_2011}, 
implement various approximations of the complex physics at work to compute stellar characteristics 
($R_*, L_*, g, T_\mathrm{eff}$)  as a function of input parameters. 
The main stellar parameters are mass $M_*$ and age, and the initial elemental composition (metallicity). 
After the initial models, the development of early stellar evolution models somewhat stagnated, 
largely because the models were not challenged by sufficiently accurate observations. 
Recent unexpected discoveries, such as that of the ``twin'' binary star Par 1802 \citep{Stassun+etal_2008}, 
have raised new questions and fostered new developments. Additional complexity such as stellar rotation 
\citep[e.g.][]{Maeder+Meynet_2005}, magnetic fields \citep[e.g.][]{Macdonald+Mullan_2010, Morales+etal_2010}, 
and the history of accretion \citep{Baraffe+Chabrier_2010}, starts being incorporated in existing models.

These models must be validated by comparison between their predictions and actual observations. 
Several possibilities exist: 1/ Radius $R_*$ is  a primary characteristic, but its direct 
determination is only possible through optical/IR interferometric measurements within $\sim 25$\,pc, 
or in the specific case of eclipsing binaries, and hence not suitable for distant young stars. 
2/ Surface gravity $g$ can be derived from observed spectra, providing a proxy for the 
mass (through the $M_* \propto  L_* g T_\mathrm{eff}^{-4}$ relation), but the mass uncertainty 
then scales as $D^2$. 3/ Metal depletion, essentially that of Li 
for young stars \citep[e.g.][]{White+Hillenbrand_2005}, 
and pulsation modes from astero-seismologic measurements can also be age indicators, 
but are restricted to (very) limited ranges of age and mass for young stars. The only prime 
parameter that  can be unambiguously compared with observations remains the stellar mass.

The Keplerian rotation of disks \citep{Guilloteau+Dutrey_1998} is the only method that
can be used to measure $M_*$, or more precisely $M_*/D$, the Mass to Distance ratio for single stars.
As $L_*$ scales as $D^2$, stars can be accurately placed in a modified HR diagram: $L/M^2 \mathrm{~vs~} T_\mathrm{eff}$,
thereby canceling the impact of the distance uncertainty (that can be large for the star formation regions).
Our pioneering work  using this simple method indeed suggested that some of the available evolutionary models
did not agree with these direct mass determinations \citep{Simon+etal_2000}.
It was based on only 8 stars, but little progress has been made since. For isolated sources, 
CQ Tau was measured by \citet{Chapillon+etal_2008}, and MWC 758 by \citet{Isella+etal_2010a} 
using CO \citep[but both stars suffer from large distance uncertainty, see][]{Chapillon+etal_2008}, while the mass of HH\,30
(unfortunately a binary) was derived by \citet{Pety+etal_2006} from $^{13}$CO.
For embedded sources, CO disk detections were reported by \citet{Schaefer+etal_2009} for LkHa 358, 
GO Tau, Haro 6-13 and Haro 6-33, and by \citet{Guilloteau+etal_2011} for FT Tau, but no  accurate masses could be
derived because of contamination of the disk emission by emission or absorption from 
their surrounding environments (clouds, envelopes and/or outflows) or from molecular clouds along the line of sight.
The effectiveness of dynamical mass measurements is attested by the work of \citet{Rosenfeld+etal_2012},
who compared the dynamical mass derived from CO observations of the circumbinary disk of
VX 4046 Sgr to the mass obtained from the analysis of the radial velocity curves of this spectroscopic binary.

\citet{Pietu+etal_2007} and \citet{Dutrey+etal_2008} improved the results
on DM Tau, LkCa\,15, MWC\,480, and GM Aur using CO isotopologues, and these 4 stars
remain the only single young low mass stars with accurate masses.

Beating the contamination problem is a pre-requisite to determine accurate dynamical masses. In \citet{Guilloteau+etal_2013},
we showed through a survey of 40 stars that CN N=2-1 transition is a good tracer for this purpose.
It appears in general free of contamination from clouds, and is strong enough in many disks to be a sensitive
tracer of the disk kinematics. We use here this property to study a sample of 12 stars in CN N=2-1 using
high angular and spectral resolution spectro-imaging with the IRAM Plateau de Bure interferometer, and
derive accurate masses for 11 of them.

\section{Observations and Analysis}
\label{sec:obs}

\subsection{Source Sample}
\label{sec:sub:sample}
Our sample is derived from the study of \citet{Guilloteau+etal_2013}. It
includes all ``bona-fide'' disks with strong enough CN emission to be imaged
in a short (4 hours per source) time with the IRAM interferometer. Sources
exhibiting potential contamination from outflows or envelopes were deliberately excluded
at this stage, with the exception of the enigmatic embedded object DG Tau B.

Our sample contains 12 stars: 9 T Tauri stars, one HAe (MWC\,480) and two
embedded objects, IRAS04302+2247 (the Butterfly star) and DG Tau B. All
stars are single, except HV Tau, which is a triple system.

Data for the well known, isolated (from any surrounding cloud), objects like
DM Tau, LkCa 15 and MWC\,480 are taken from \citet{Chapillon+etal_2012}. These
sources have been observed in many other molecular
lines, and thus serve as a probe of the reliability of CN as a dynamical
mass tracer.

Newly observed sources include objects for which previous
attempts to derive dynamical masses were affected by low S/N and contamination from
molecular clouds: CY Tau, DL Tau by \citep{Simon+etal_2000}, and GO Tau by \citet{Schaefer+etal_2009}.
The remaining sources had no previous dynamical mass measurements: DN Tau and IQ Tau
(which had been observed in $^{12}$CO by \citet{Schaefer+etal_2009} but not detected, 
presumably because of contamination by a molecular cloud), HV Tau, IRAS04302+2247 and DG Tau B.

\subsection{Observations}
\label{sec:sub:obs}
All observations were carried out with the IRAM interferometer.
\object{DM Tau}, \object{LkCa 15} and \object{MWC 480} have been reported by \citet{Chapillon+etal_2012}.
The characteristic angular resolution is $1.6 \times 1.0''$ for DM Tau, and
$1.3 \times 0.8''$ for MWC\,480 and LkCa 15 ($1.0 \times 0.65''$ with uniform weighting)
which have baselines as long as 330 m (250 k$\lambda$).

\object{CI Tau}, \object{CY Tau} and \object{GO Tau} were observed on the night of Nov 6 to 7, 2010 in C configuration in a track sharing mode.
The single-sideband, dual polarization receivers were tuned to cover
the CN N=2-1 transition around 226.784 GHz.
This transition has 19 hyperfine components, with relative intensities
spanning 2 orders of magnitude.
The high resolution backend covered the 6 strongest hyperfine components (which account for 
81.25 \% of the total line intensity, see Table \ref{tab:lines}), with a channel separation
of 39 kHz (0.052 km/s at this frequency), and an effective spectral resolution
about 1.6 times coarser given the apodization applied in the correlator.
The effective integration time is about
2 hours per source, leading to an rms noise about 20 mJy/beam, or 0.3 K
after resampling at 0.206 km/s spectral resolution.
In addition, the wideband correlator provided a coverage of 4 GHz in each
polarization.

With a longest baseline about 180 m (130 k$\lambda$), the C configuration provides an effective
angular resolution of $1.3 \times 1.0''$
at PA near 30$^\circ$ (with a slight dependency on exact $UV$ coverage).

\begin{table}[!h]
\caption{Frequencies of observed CN N=2-1 transitions}
\begin{tabular}{lcr}
 Frequency &  Hyperfine & Relative \\
 (MHz)  &  Transition     &  Intensity \\
 \hline
 %
 %
 226659.5584 &  J=3/2-1/2, F=5/2-3/2 & 0.1667 \\
 226663.6928 &  J=3/2-1/2, F=1/2-1/2 & 0.0494 \\
 226679.3114 &  J=3/2-1/2, F=3/2-1/2 & 0.0617 \\
 226874.1908 &  J=5/2-3/2, F=5/2-3/2 & 0.1680 \\
 226874.7813 &  J=5/2-3/2, F=7/2-5/2 & 0.2667 \\
 226875.8960 &  J=5/2-3/2, F=3/2-1/2 & 0.1000 \\
 %
\hline
\end{tabular}
\tablefoot{CN N=2-1 line frequencies were measured in laboratory by \citet{Skatrud+etal_1983};
we use here the fitted values from the CDMS Database \citep{CDMS_2001}.}
\label{tab:lines}
\end{table}

The same instrumental setting and antenna configuration were also used for the 6 other
sources, which were observed on 3 contiguous nights from Nov 19, 2012 to
Nov 22, 2012, two sources at a time.
The effective integration time is about 3 hours per source, the rms noise
35 mJy/beam, or 0.6 K at the full spectral resolution. Phase noise
were 20 to $50^\circ$ on \object{DL Tau} and \object{HV Tau}, 15 to $40^\circ$ on \object{DN Tau}
and \object{IRAS042302+2247}, and 20 to $60^\circ$ on \object{IQ Tau} and \object{DG Tau B}.

All data was calibrated using the CLIC program in the GILDAS package. Bandpass
calibration was made on strong quasars (3C84 or 3C454.3). Standard phase and
amplitude calibrations were made using nearby quasars 0400+258 and 0507+179.
Both quasars were weakly polarized, and the measured polarization information
was taken into account in the amplitude calibration process.
The flux calibration is based on a model flux for MWC 349 of 1.96 Jy at this
frequency. The repeatability was better than 3\% for the 3 consecutive nights where
the same quasars were used.

In the continuum, the expected thermal noise is around 0.4 -- 0.6 mJy/beam for the
newly observed sources. However, the original images are dynamic range limited.
As the sources are compact and strong, phase-only self calibration was used to improve the
on-source phase noise.  This brought the final noise to within a factor 1.5--2 of
the expected value. The spectral line data is essentially noise limited,
 so that the application of the self-calibration solution does not significantly
 affect the results. Thus, for CN, the reported values only use the original
 (non self-calibrated) data set, with the exception of CY Tau, which will
 be discussed in more details.

\subsection{Analysis Method}
The continuum data were fit by a simple 2-D elliptical Gaussian model. Position
angle of the dust disk major axis and inclination are reported in Table
\ref{tab:geom}. Dust disk inclinations
are derived from minor to major axis ratio. For highly inclined objects, they thus
underestimate the true inclination, because of the flared disk geometry.

For the spectral line data, we
use the DiskFit tool \citep{Pietu+etal_2007} to fit a parametric
disk model to the calibrated visibilities. The disk model assumes
power laws for all major quantities: CN column density, CN rotation
temperature, CN scale height distribution, as well as for the
rotation velocity. Power laws are expressed in the from
\begin{equation}
    F(r) = F_{100} \left(r/100\,\mathrm{AU}\right)^{-f}
\end{equation}
A single model has a priori
16 free parameters:\\
- five geometric parameters: position $x_0,y_0$, rotation axis position angle $PA$, inclination $i$
and systemic velocity $V_\mathrm{LSR}$\\
- two parameters (value at 100 AU and exponent) for each power law: temperature
$T_{100}, q$, surface density $\Sigma_{100}, p$, rotation velocity $V_{100}, v$ and
scale height $H_{100}, h$\\
- the inner and outer radius, R$_\mathrm{int}$ and R$_\mathrm{out}$\\
- and finally, the local linewidth $d V$, which includes the thermal and
turbulent broadening.

For Keplerian rotation, we expect an exponent $v = 0.50$.  We report instead
the departure from Keplerian rotation, $\delta v$ defined as $v = 0.50 + 0.01 \delta v$.
We neglect any radial dependency of the local linewidth.
For the orientation, we follow the convention presented in \citet{Pietu+etal_2007}
by giving the PA of the rotation axis oriented by the disk rotation, which is thus
defined between 0 and 360$^\circ$.

Among these 16 parameters, some have negligible influence, such as the inner radius, R$_\mathrm{int}$,
which can in general be set to an arbitrary low value $< 20$ AU. 
Others may be too strongly coupled together
to be separately derived, such as those controlling the column density and temperature
profiles. On the contrary, the position ($x_0,y_0$), which is mostly determined by the phases, is
essentially completely decoupled from the other parameters which are determined
by the amplitude of the visibilities. In practice, the only strong coupling which matters
for our objective (the stellar mass measurement) is between $V_{100}$ and $\sin(i)$.
The scale height parameters, $H_{100}$ and $h$, will be discussed more thoroughly in
Sec.\ref{sec:scale}. 
For a given velocity resolution, the mass precision depends, to first
order, on the product of the signal to noise ratio of the line brightness
and the ratio of disk size to angular resolution, provided that each
of these is substantially greater than 1. The velocity resolution must be sufficient to sample the local
line width $dV$, which is typically 0.2 -- 0.4 km.s$^{-1}$; 
undersampling this width would degrade the method precision, 
but using higher spectral resolutions provide no further improvement.

Minimization is performed using a modified Levenberg-Marquardt method, with
multiple restarts to avoid being trapped in local minima.
Error bars were computed from the covariance matrix. They thus should be
interpreted with some caution in case of non-Gaussian distributions. However,
for $V_{100}$, the problem is well behaved and the covariance matrix provides
a good estimate of its error, provided the inclination $i$ is moderate ($25-30^\circ < i < 80^\circ$, 
so that the error on $i$ remains to first order symmetric).
For the two sources which have extreme inclinations, CY Tau and HV Tau, 
we used a different procedure to derive the uncertainty
on the inclination: we instead computed the $\chi^2$ curve as a function
of $i$, by minimizing over all other parameters.

\citet{Pietu+etal_2007} present a thorough discussion of the merit of fitting
continuum subtracted data or fitting the continuum together with the spectral line.
As the observed transitions are essentially optically thin, we use here continuum
subtracted data.

All sources were analyzed in the same way, except \object{DG Tau B} (see Sec.\ref{sec:dgtaub}).

\begin{table}
\caption{Geometric parameters }
{\tiny
\begin{tabular}{l|r|r|r|r}
\hline
 Source &  \multicolumn{2}{c|}{Orientation ($^\circ$)} & \multicolumn{2}{c}{Inclination ($^\circ$)} \\
 name & CN & Cont.  & CN  & Cont. \\
\hline
CI Tau &     281.5 $\pm$      0.5 &    282.7 $\pm$      1.1 &     51.0 $\pm$      1.9 &     45.7 $\pm$      1.1 \\
CY Tau &      62.5 $\pm$      1.8 &     35.6 $\pm$      6.2 &     24.4 $\pm$      2.4 &     34.4 $\pm$      9.0 \\
GO Tau &     111.1 $\pm$      0.3 &    114.8 $\pm$      2.3 &     54.5 $\pm$      0.5 &     48.6 $\pm$      2.6 \\
HV Tau C &   197.8 $\pm$      0.9 &    199.5 $\pm$      0.8 &     89.1 $\pm$      3.0 &     75.8 $\pm$      1.1 \\
DL Tau &     322.6 $\pm$      0.6 &    320.5 $\pm$      0.3 &     43.6 $\pm$      2.5 &     42.3 $\pm$      0.3 \\
IQ Tau &     309.9 $\pm$      1.0 &    313.8 $\pm$      1.1 &     57.9 $\pm$      6.7 &     58.2 $\pm$      1.6 \\
DG Tau B &                        &     25.7 $\pm$      0.3 &                         &     58.0 $\pm$      0.4 \\
DN Tau &     174.2 $\pm$      2.2 &    185.8 $\pm$      5.7 &     29.7 $\pm$      2.7 &     26.8 $\pm$      3.9 \\
04302+2247 & 297.6 $\pm$      2.1 &    254.2 $\pm$      1.4 &     58.9 $\pm$      2.1 &     78.1 $\pm$      3.0 \\
\hline
\end{tabular}
}
\label{tab:geom}
\tablefoot{Geometric parameters derived from CN and 1.3 mm continuum data.
Orientation and inclinations are those of the rotation axis,
following the convention of \citet{Pietu+etal_2007}.}
\end{table}

\subsection{CO Data}
\label{sec:codata}

We also re-analyzed CO J=2-1 data from \citep{Simon+etal_2000}, completed with
the high angular resolution data obtained during the continuum survey of \citet{Guilloteau+etal_2011},
for 3 sources: CI Tau, CY Tau and DL Tau. All sources suffer from contamination. Based on inspection of the images,
we avoided the velocity range $[4.20,7.20] \kms$ for DL Tau \citep[see also Fig.B.22 in][for the
contaminated range in $^{13}$CO]{Guilloteau+etal_2013}, the range $[5.50,7.25] \kms$ for CY Tau, and  $[3.90,6.50] \kms$ for CI Tau.  
This masking procedure makes the derivation of the systemic velocity more uncertain:
formal errors are not reliable for this parameter because
of the bias introduced by the channel selection. The possible bias on the systemic velocity also
affects the fitted rotation velocity, but the resulting bias is not included in the formal error. Disk size and inclination may
also be biased if the masked velocity range is large and encompasses the systemic velocity.

Comparison with the CN results is given in Table \ref{tab:co-cn}.
Unlike the CN data, the CO results are dominated by high resolution data, but both agree
within the noise.  The agreement also applies for the disk size: although CO and CN may have
different radial distributions, they have identical outer radii, $\sim 460$ AU for DL Tau and
$\sim 280$ AU for CY Tau (with typical formal errors about 15 AU).

\begin{table}
\caption{Comparison between CO and CN results.}
\label{tab:co-cn}
{\tiny 
\begin{tabular}{l|r|r|r|r}
\hline
Tracer &  Orientation  & Inclination & V$_\mathrm{LSR}$ & V$_{100}$ sin(i) \\
       &  ($^\circ$) & ($^\circ$) & ($\kms$) & ($\kms$)  \\
\hline
\multicolumn{5}{c}{CY Tau} \\
CN &     62.5 $\pm$ 1.8 & $24.0 \pm 2.4$ & $7.26 \pm 0.01$  &  $0.99 \pm 0.06$ \\ 
Cont. (a) & $36 \pm 6$  & $34 \pm 9$ &  \multicolumn{2}{c}{--} \\
CO        & $63 \pm 1$  & $29 \pm 5$ & $7.27 \pm 0.02$ & $0.95 \pm 0.05$ \\
Cont. (b) & $63 \pm 5$  & $34 \pm 3$ &  \multicolumn{2}{c}{--} \\
\hline
\multicolumn{5}{c}{DL Tau} \\
CN &     322.6 $\pm$ 0.6 & $43.6 \pm 2.5$ & $6.10 \pm 0.01$  &  $1.94 \pm 0.02$ \\
Cont. (a) & $320.5 \pm 0.3$  & $42.3 \pm 0.3$ &  \multicolumn{2}{c}{--} \\
CO    (*) & $321.0 \pm 2.4$  & $39.6 \pm 1.3$ & $6.00 \pm 0.10$ & $2.04 \pm 0.10$ \\
Cont. (b) & $321 \pm 3$  & $38 \pm 2$ &  \multicolumn{2}{c}{--} \\
\hline
\multicolumn{5}{c}{CI Tau} \\
CN &     281.5 $\pm$ 0.5 & $50.1 \pm 1.9 $ & $5.73 \pm 0.02$  &  $2.61 \pm 0.04$ \\
Cont. (a) & $282.7 \pm 1.1$  & $45.7 \pm 1.1$ &  \multicolumn{2}{c}{--} \\
CO        & $285.2 \pm 0.8$  & $53.3 \pm 1.9$ & $5.77 \pm 0.03$ & $2.46 \pm 0.05$ \\
Cont. (b) & $286.0 \pm 2.1$  & $53.8 \pm 1.7$ &  \multicolumn{2}{c}{--} \\
\hline
\end{tabular}
}
\tablefoot{(a) This work with 1.3$''$ resolution. (b) From \citet{Guilloteau+etal_2011}
with $\simeq 0.5''$ resolution. (*) \citet{Guilloteau+etal_2011} incorrectly
reported the orientation modulo $180^\circ$ for this source.}
\end{table}

\section{Results and method limitations}

\begin{figure*}
\includegraphics[width=18.0cm]{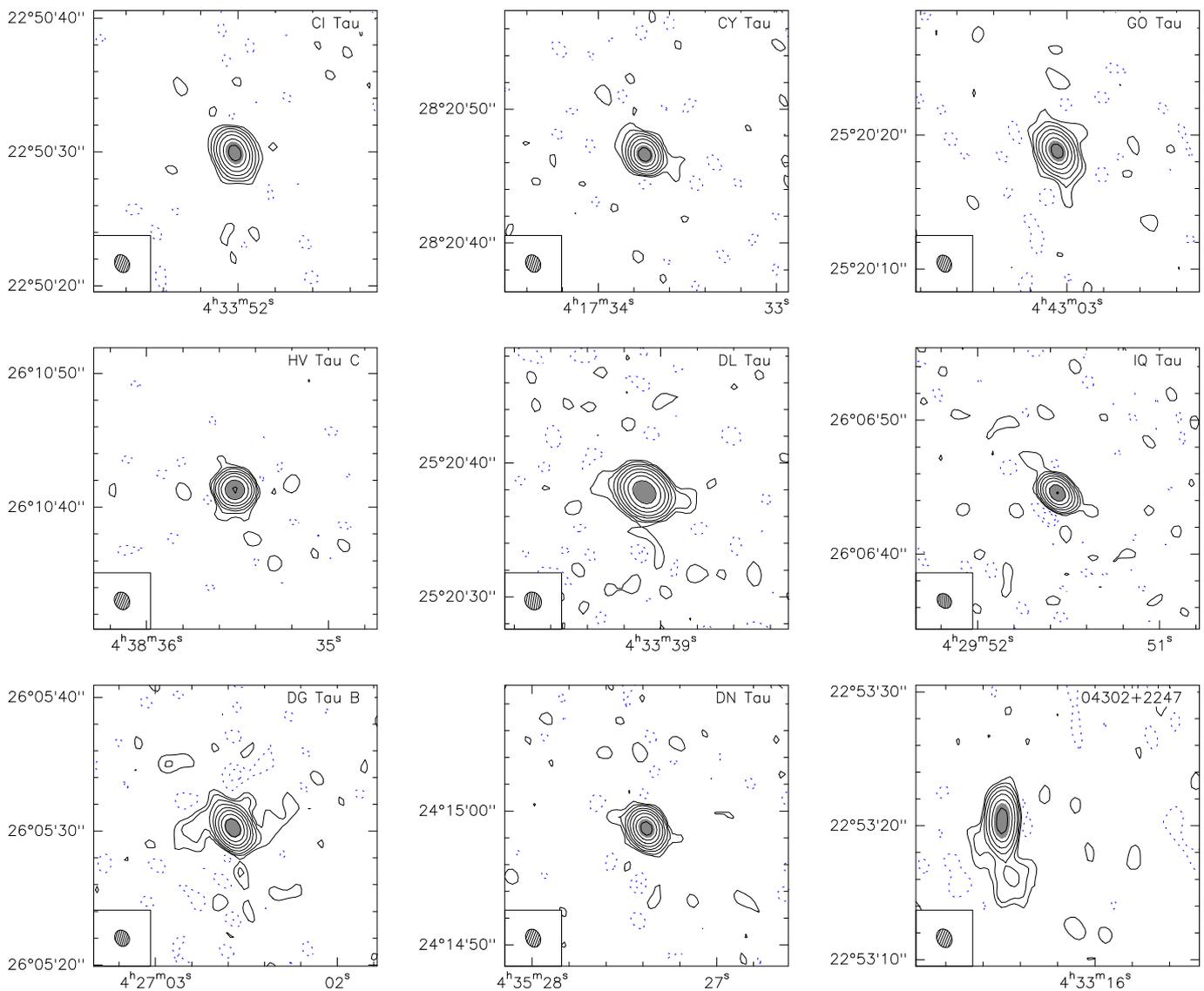}
\caption{Continuum images for all newly observed sources.
Contour levels are logarithmically spaced, by a factor 2: -2 and 2, 4, 8, 16, 32, 64, 128 and 256 $\sigma$.
Rms noise (in mJy/beam) and peak signal to noise are: CI Tau 0.6 and 180, CY Tau 0.5 and 190,
GO Tau 0.26 and 180, HV Tau 0.12 and 230, DL Tau 0.23 and 550, IQ Tau 0.23 and 270,
DG Tau B 0.7 (dynamic range limited) and 450, DN Tau 0.25 and 360, and IRAS04302+2247 0.4 and 180
}
\label{fig:continuum}
\end{figure*}

Figure \ref{fig:continuum} presents the self-calibrated continuum images.

For each source, we have two or three data cubes of $128 \times 128$ pixels
and 460 channels each. Signal is spread over 50 to 100 channels, but the signal
to noise per channel is in general rather low (see for example Fig.\ref{fig:dn_tau}).
It is hopeless to present the full data cubes. For each source, we present instead two
``optimal'' quantities.
The first quantity is the set of spectra for each (group of) hyperfine
component, integrated over the disk area defined below.
The second one are images of optimally filtered spectra,
as computed for N$_2$H$^+$ by \citet{Dutrey+etal_2007}.
Each channel is multiplied by the intensity of the integrated spectrum
predicted by the best fit model. All channels are then summed together:
the resulting quantity has the dimension of an intensity squared summed 
over velocity, and no simple
physical interpretation, but gives the signal-to-noise
for detecting line emission which matches the model profile. 
This signal to noise image shows where the emission is located.
The process is applied  separately for
each (group of) hyperfine component, and then globally.
The global S/N map serves as a mask to compute the integrated
spectra, using a 2 $\sigma$ threshold.

\begin{figure*} 
\includegraphics[width=18.0cm]{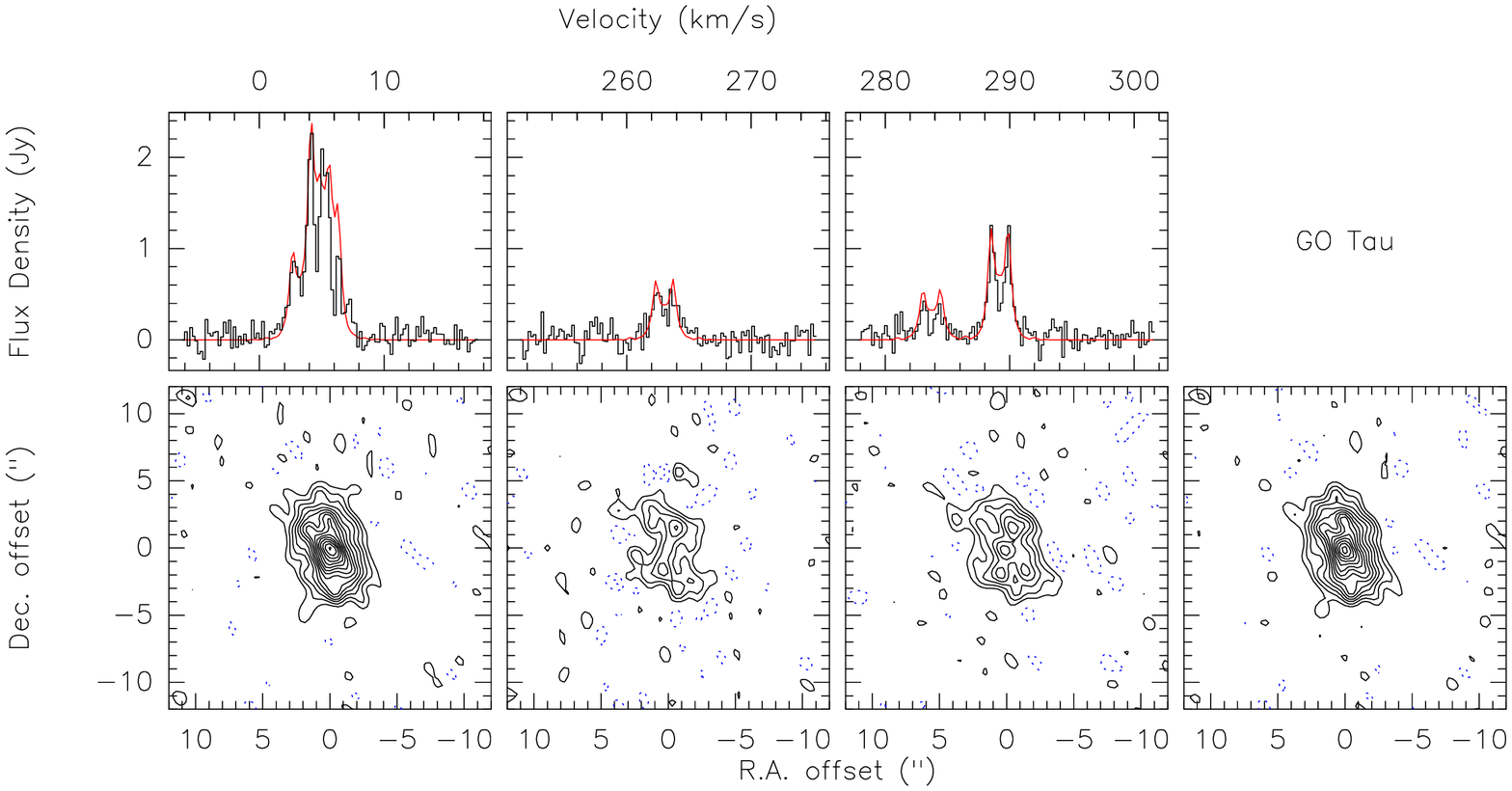}
\caption{Results for GO Tau. Top: Integrated line flux for CN N=2-1 hyperfine components
(histogram) with the best fit profile superimposed (red curve).
Bottom: Signal to noise maps for each group of component, and
for all observed components (rightmost panel); contour spacing is $2 \sigma$.
}
\label{fig:nice}
\end{figure*}

An example of these integrated spectra and signal to noise maps is
given in Fig.\ref{fig:nice}. All other sources are shown
in Appendix \ref{app:figures} in Figs.\ref{fig:dm_tau}-\ref{fig:go_tau}, except for DG Tau B
which is discussed in Sect.\ref{sec:dgtaub}.
Unless noted, the spectra have been smoothed
to 0.206 $\kms$ resolution for better clarity. The agreement between
the observed line profiles and the best fit results appears sometimes limited, but
this is a result of the difficulty to deconvolve low signal to noise data
combined with synthesized beams with substantial sidelobes. Under such circumstances,
the deconvolution cannot recover the total flux. 

\begin{figure}[!h] 
\includegraphics[width=6.0cm]{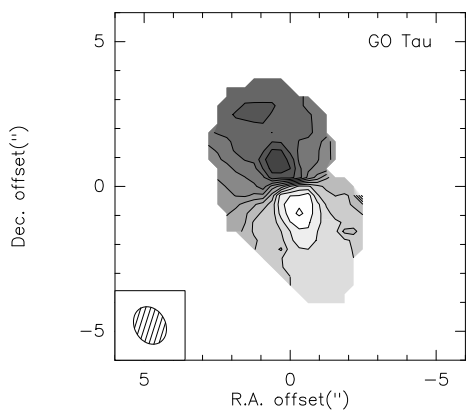}
\caption{Velocity gradient for GO Tau. Contour spacing is 0.2 km\,s$^{-1}$.
}
\label{fig:gradient}
\end{figure}

All sources show clear evidence for rotation, but illustrating the velocity gradient 
is not straightforward because of the multiple 
(and for the strongest, blended) hyperfine components (see Fig.\ref{fig:quality}). 
We show in Fig.\ref{fig:gradient}
the first moment map derived from the isolated hfs component near 290 km\,s$^{-1}$ in GO Tau.
The fitting procedure, which takes all hyperfine components into account, retrieves
the related information (orientation, velocity field and inclination) with much better precision.

Relevant parameters from the fit results are presented in Table \ref{tab:vmass}.
To illustrate the fit quality, we present in Appendix (Fig.\ref{fig:quality}) channel
maps for the strongest group of hyperfine components in GO Tau: observations, best
model and residuals. On average, there is no systematic dependency of the
residuals with velocities, and channels with strong emission have similar
residuals than channels without. Although a few channels have some systematic residuals, 
this can be ascribed by the limitations of our disk model, such as the assumption
of power law for the CN surface density, and should not affect the derived
velocity field.  For all other sources, we obtained significantly less signal to noise than
for GO Tau, and noise is the limiting factor in the velocity field derivation.

A comparison between geometric parameters derived from CN and dust emission is given in
Table \ref{tab:geom}.
The stellar mass is derived from the Keplerian rotation of the disks. Because the maps provide only
an angular scale the derived masses are proportional to the distance
to the star.  The masses listed in Table \ref{tab:vmass} are given at the
star-forming region's (SFR) average distance, 140 pc \citep{Kenyon+etal_1994}.

\begin{table*}
\caption{Disk and Star parameters derived from CN}
\begin{tabular}{l|c|c|c|c|c}
\hline
 Source &  V$_{100}$ & i & $R_\mathrm{out}$ & M$_*$  & $\delta v$ \\
 name &  (km\,s$^{-1}$) & ($^\circ$) & AU & ($\Msun$) & \\
\hline
DM Tau &      2.31 $\pm$     0.17 &    -30.9 $\pm$      2.9 &    641 $\pm$     19 &     0.60 $\pm$     0.09 &    0.1 $\pm$    1.5 \\
MWC 480 &      4.03 $\pm$     0.41 &     36.3 $\pm$      2.5 &    539 $\pm$     39 &     1.83 $\pm$     0.37 &   -2.2 $\pm$    2.0 \\
LkCa 15 &      3.11 $\pm$     0.10 &     47.0 $\pm$      1.3 &    567 $\pm$     39 &     1.09 $\pm$     0.07 &    0.0 $\pm$    2.4 \\
CI Tau &      2.67 $\pm$     0.03 &     51.0 $\pm$      0.9 &    520 $\pm$     13 &     0.80 $\pm$     0.02 &    -2.7 $\pm$    2.0 \\
CY Tau &      2.36 $\pm$     0.12 &     24.0 $\pm$      2.0 &    295 $\pm$     11 &     0.63 $\pm$     0.05 &    1.7 $\pm$    1.7 \\ 
GO Tau &      2.07 $\pm$     0.01 &     54.5 $\pm$      0.5 &    587 $\pm$     55 &     0.48 $\pm$     0.01 &    4.0 $\pm$    2.0 \\ 
HV Tau C &      3.76 $\pm$     0.10 &     89.1 $\pm$      3.0 &    256 $\pm$     51 &     1.59 $\pm$     0.08 &   -0.0 $\pm$    2.9 \\
DL Tau &      2.83 $\pm$     0.04 &     44.1 $\pm$      2.6 &    463 $\pm$      6 &     0.91 $\pm$     0.02 &    1.9 $\pm$    1.1 \\ 
IQ Tau &      2.64 $\pm$     0.02 &     56.3 $\pm$      3.9 &    225 $\pm$     21 &     0.79 $\pm$     0.02 &   -0.3 $\pm$    4.9 \\ 
DN Tau &      2.91 $\pm$     0.25 &     29.2 $\pm$      3.0 &    241 $\pm$      7 &     0.95 $\pm$     0.16 &   -0.6 $\pm$    1.8 \\ 
04302+2247 &      4.18 $\pm$     0.09 &     58.9 $\pm$      2.1 &    750 $\pm$     56 &     1.97 $\pm$     0.08 &   -0.4 $\pm$    2.3 \\ 
\hline
\end{tabular}
\label{tab:vmass}
\tablefoot{$\delta v$ is the departure from Keplerian rotation:
$v(r) \propto r^{-(0.50+0.01 \,\delta v)}$. }
\end{table*}

\subsection{CN as a dynamical mass tracer}

A first important result from this study is an unambiguous confirmation that CN
is essentially unaffected by contamination from the molecular clouds. The full kinematic
pattern of the disk is visible, leading to accurate determination of the
systemic velocity. However, the disk interpretation does not apply for the
two embedded (presumably younger) objects DG Tau B and IRAS04302+2247.

A second essential result from Table \ref{tab:vmass} is that all sources appear
in Keplerian rotation. The weighted mean deviation from the Keplerian exponent
$v = 0.50 + 0.01 \delta v$ is $\delta v = 0.5 \pm 0.6$.
Thus, we can safely interpret the rotation pattern as being driven by a central
mass.

The third result is the good agreement between geometric parameters (position angle
and inclination) derived from other tracers. This is shown for CI Tau, DL Tau and
CY Tau in Table \ref{tab:co-cn}.
This agreement is important, because the geometric parameters are affected by
different systematic effects due to calibration uncertainties.
For continuum data, phase errors \citep[for the high
resolution data of][]{Guilloteau+etal_2011} or amplitude errors (for our self-calibrated
data) can bias the result beyond the thermal noise.  For spectral line data, the
orientation is defined by the velocity gradient, and thus depends on the
phase bandpass calibration.

Finally, in previously studied isolated sources,
the systemic velocity derived from CN is (within the noise induced uncertainties) in
agreement with values derived from other molecular transitions \citep[see][for DM Tau,
LkCa 15 and MWC 480]{Pietu+etal_2007}.

We thus conclude that, within the uncertainties of the measurements reported here, 
CN is a good tracer of the dynamical mass. This dynamical mass may however overestimate 
the stellar mass: given the angular scale of the study, it includes any contribution 
from any compact ($< 20-30$ AU radius)  disk that may surround
the star.


\subsection{The vertical distribution of CN}
\label{sec:scale}

In our analysis, CN is assumed to be homogeneously distributed in the vertical direction,
i.e. to be distributed vertically following a gaussian whose characteristic size is the 
disk scale height.
The scale height was initially arbitrarily
fixed to $H_{100} = 16.5$ AU and $h=-1.25$ (mildly flared disk). However, we expect
CN to be located at the top of the molecular layer, between the highly irradiated surface
layer where molecules are photo-dissociated and the colder disk plane where molecules
condense on grains. Although this expected distribution has so far not been confirmed
by imaging studies, and even faces some difficulty when considering the expected temperature
in this molecular layer \citep{Chapillon+etal_2012}, it is important to evaluate if this
can affect the derived disk inclination and hence, the stellar mass measurement.

We did that in 3 steps. First, we explored different disk thickness.
This did not significantly affect the derived dynamical mass.
Second, we
treated the scale height as a free parameter. We found
in general larger values for $H_{100}$, and flatter disks $h \approx -1.0$, which
is consistent with the molecules being closer to the disk surface. However, the
number of free parameters becomes large, and the fits sometimes converge towards
unrealistic solutions (e.g. very large $H_{100}$ and $h > -1.0$).
Third, we used the method described by \citet{Guilloteau+etal_2012} for
the analysis of  CS in DM Tau. We assumed CN molecules to be absent at any
point where the H$_2$ column disk towards the disk surface $\Sigma_o(r,z)$ is larger than
a given value, $\Sigma_d$.  $\Sigma_o(r,z)$ density is computed using the prescribed scale 
height ($H_{100} = 16.5$ AU and $h = -1.25$), and with the surface density profile
corresponding to the best fit viscous disk model to the 226 GHz continuum
data:
\begin{equation}\label{eq:edge}
\Sigma_g(r) = \Sigma_0 \left(\frac{r}{R_0}\right)^{-\gamma}  \exp\left(-(r/R_c)^{2-\gamma}\right) .
\end{equation}
Despite the much more limited angular resolution of the new data, we reached
a similar precision on $\gamma$ and $R_c$ as \citet{Guilloteau+etal_2011}, thanks to
the higher sensitivity and lower phase noise due to self-calibration. Furthermore,
the values found
for $\gamma$ and $R_c$ were within the errors equal to those measured by \citet{Guilloteau+etal_2011}.
We varied the depletion scale height $\Sigma_d$ between $10^{21}$ and $10^{24}$
cm$^{-2}$. The most extreme values provided significantly worse fit to the data,
with a best fit $\Sigma_d$ around $10^{23}$ found for most sources. The derived disk
inclinations and dynamical masses only change by
small amounts as a function of $\Sigma_d$: less than $\sim \sigma/3$ over the acceptable
range of values for $\Sigma_d$, and even at most about $1 \sigma$ for more extreme
values.

We thus conclude that the uncertainties in the spatial distribution of CN due
to our limited knowledge of the disk structure and chemistry do not affect the reliability
of the CN N=2-1 transition as a tracer of the dynamical mass.

\subsection{Special sources}

We derived accurate dynamical masses for most sources in our sample.
However, a few sources are peculiar in this respect, because of unfavorable 
inclination and orientation,
evolutionary status or multiplicity. We discuss here these sources 
which we omit in further analysis.

\subsubsection{HV Tau}
\label{sec:hvtau}
HV Tau is a triple system: AB is a close visual binary with angular separation around
70 mas \citep{Simon+etal_1996}, and HV Tau C is a much fainter T Tauri star located 4$''$ NE of AB.
A nearly edge-on compact dust disk surrounds HV Tau C \citep{Monin+Bouvier_2000}, but AB shows no
IR excess and no sign of accretion \citep{White+Ghez_2001}. The system was modeled in detail by
\citet{Duchene+etal_2010}, who showed that mm-emission only comes from the HV Tau C circumstellar
disk, which also exhibit $^{12}$CO emission compatible with Keplerian rotation, although only
the line wings are clearly visible.

CN emission was detected using the IRAM 30-m by \citet{Guilloteau+etal_2013}, who attributed it
to HV Tau C based on
the previous non-detection of any circumstellar material around AB by \citet{Duchene+etal_2010}.
Our images confirm this association. The derived dynamical mass, $\sim 1.6 \Msun$, is much larger
than suggested by apparent spectral type \citep[K6,][]{White+Hillenbrand_2004}. Given the
system complexity, a more complete study of the HV Tau multiple system will be presented in
a separate paper.

With IQ Tau, HV Tau C is another good example of disk detection through CN while CO emission 
is heavily contaminated by the molecular cloud. Note that both stars are close to each 
other, and affected by the same molecular cloud. The higher inclination of HV Tau C allows 
the line wings to remain visible in CO.

\subsubsection{CY Tau}
\label{sec:sub:cytau}

CY Tau and DN Tau are the two sources which combine a small (280 AU) disk with a low inclination
(25-30$^\circ$). The DN Tau disk is favorably oriented with respect to the synthesized
beam, allowing the inclination to be precisely measured. On the contrary,
for CY Tau, the synthesized beam major axis is within $30^\circ$
of the disk rotation axis, so that the limited angular resolution in this direction makes the
inclination, and consequently the deprojected rotation velocity,  rather uncertain.  
Without self calibration for CN, the $\chi^2$ curve as a function of inclination for CY Tau has two
minima, at $\simeq 21^\circ$
and $\simeq 30^\circ$, separated by about $3 \sigma$. After application of the self-calibration
solution from the continuum data, this CN $\chi^2$ curve now exhibits a single minimum, indicating
an inclination of $\simeq 24 \pm 2^\circ$. However, low inclinations are
not strictly excluded, and the only safe conclusion is $i < 28^\circ$ at the $3 \sigma$ level, which
implies $M_* > 0.45 \Msun$ from CN only. From the CN $\chi^2$ curve, we derive $M_* = 0.63^{+0.11}_{-0.06} \Msun$.
However, using all other available measurements, the weighted mean of the inclinations is $27 \pm 2^\circ$, and
the weighted $v \sin{i} = 0.95 \pm 0.04 \kms$, which suggest a somewhat lower value for the
most likely mass, $\sim 0.51 \Msun$.

In our nominal solution from CN, we find an orientation which agree
with previous higher resolution measurements made from continuum and CO
\citep{Guilloteau+etal_2011}, see Table \ref{tab:co-cn}.
The orientation derived from the (self-calibrated) continuum is different (although marginally,
i.e. at the $2 \sigma$ level, the errors being large). Such a difference can be due to inaccuracies
in the amplitude calibration as a function of time (and hence hour angle).
Similar calibration inaccuracies cannot significantly affect
other disks, since their axis ratio is much larger than for CY Tau
(or better constrained because of a more favorable orientation in the case of DN Tau).

\subsubsection{The Butterfly star, IRAS\,04302+2247}

\begin{figure}
\includegraphics[width=8.5cm]{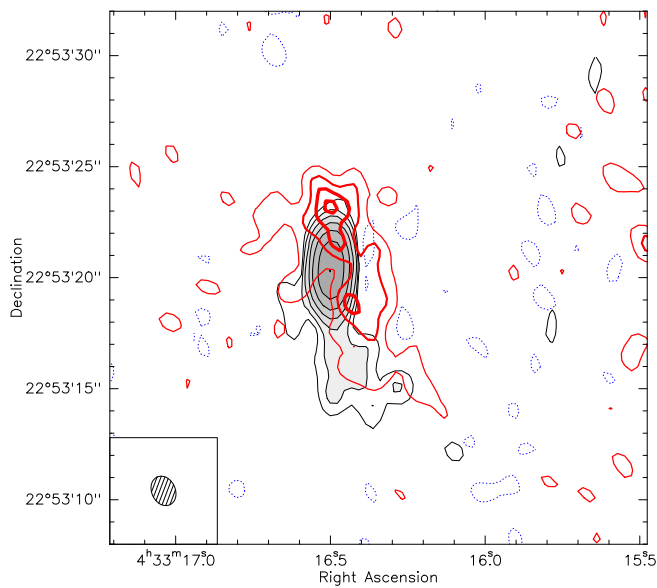}
\caption{Superposition of the CN N=2-1 emission (in red contours) over the continuum
emission (in grey scale) in IRAS04302+2247. Contour levels are 2,4,6,8 $\sigma$ for
CN, and 2, 4, 8, 16 and 32 $\sigma$ for the 1.4\,mm continuum emission.}
\label{fig:butter-c}
\end{figure}

In this embedded source, CN is clearly not peaking towards the continuum (see
Fig.\ref{fig:butter-c}), and does not originate from the inner disk traced
by dust and CO isotopologues. Although the morphology could be the trace of
asymmetric emission from the envelope surrounding the central disk, we also
fitted CN emission by a flared, rotating disk.  The derived centroid, orientation,
and inclination differ strongly from those of the continuum disk, reflecting
the apparent asymmetry. We find an inner radius around 150 AU, similar to the
large grains disk size found by  \citet{Graefe+etal_2013}, thus indicating that
CN is not present in the dense disk, but only in the outer envelope. Despite
this different morphology, the rotation
pattern still appears nicely  Keplerian, and the derived total  mass, $1.9 \Msun$,
is consistent with the sum of the star ($1.7 \Msun$, Dutrey et al.\, 2014, in prep.) 
and disk mass.

The continuum emission clearly reveals a weak, resolved, secondary source $4''$ south of
the main disk (at PA 171$^\circ$), with a total flux density of 12 mJy. The main disk
has a flux of 138 mJy. IRAS\,04302+2247 may actually be the progenitor of
a binary (or higher multiplicity) system.

\subsubsection{DG Tau B}
\label{sec:dgtaub}

\begin{figure}
\includegraphics[width=0.7\columnwidth]{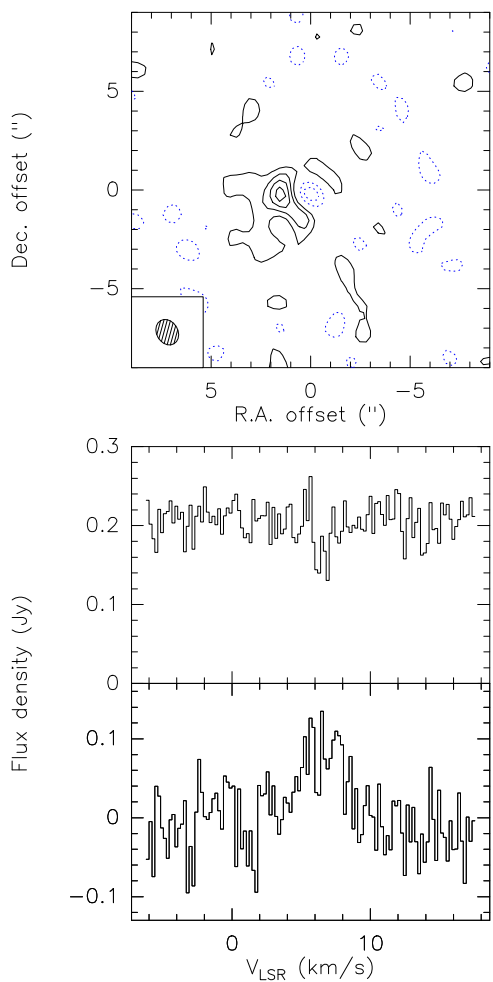}
\caption{Top: (continuum subtracted) integrated emission from CN N=2-1 main group of hyperfine components
in DG Tau B. Note the negative contours towards the continuum source at (0,0).
Middle: CN N=2-1 spectrum toward the continuum source. Bottom: integrated flux density over the emission
region.}
\label{fig:dg_tau-b}
\end{figure}

DG Tau B is another embedded object, with a powerful, one-sided, molecular outflow
\citep{Mitchell+etal_1997}.
The rather low signal to noise limits our ability to interpret the CN distribution.
However, the CN emission is not centered on the strong continuum emission. On the
contrary, we detect absorption towards the continuum source, and emission predominantly
east of the continuum peak, see Fig.\ref{fig:dg_tau-b}.   The absorption
profile is very uncertain, but could be due to a single narrow component at
$V_\mathrm{LSR} \simeq 6.2 \kms$, the hyperfine
structure leading to the impression of two velocity features.
The cloud velocity towards DG Tau B is $6.44 \kms$ from C$^{17}$O measurements
\citep{Guilloteau+etal_2013}.
The peak continuum flux is 220 mJy/beam, or about 3.7 K, and
the absorption dip has a depth of about 60 mJy/beam. Assuming the absorbing
material fills the synthesized beam, this implies an upper limit
on the excitation temperature of 7 K (taking into account the deviations
from the Rayleigh Jeans approximation), obtained for optically thick absorbing
medium. This is consistent with the expected temperatures in the surrounding
dense cloud.

On the contrary, the emission has a much larger linewidth, of order 3-4 $\kms$.
The integrated line flux is $\sim 0.6 $Jy$\kms$, about 50 \% of the flux
detected by the 30-m, which suggests that some low level emission has not been
recovered properly \footnote{This does not imply missing flux, but can be
entirely caused by the limits of deconvolution due to low signal to noise ratio}.

\begin{figure*}[ht] 
\sidecaption
\includegraphics[width=12.0cm]{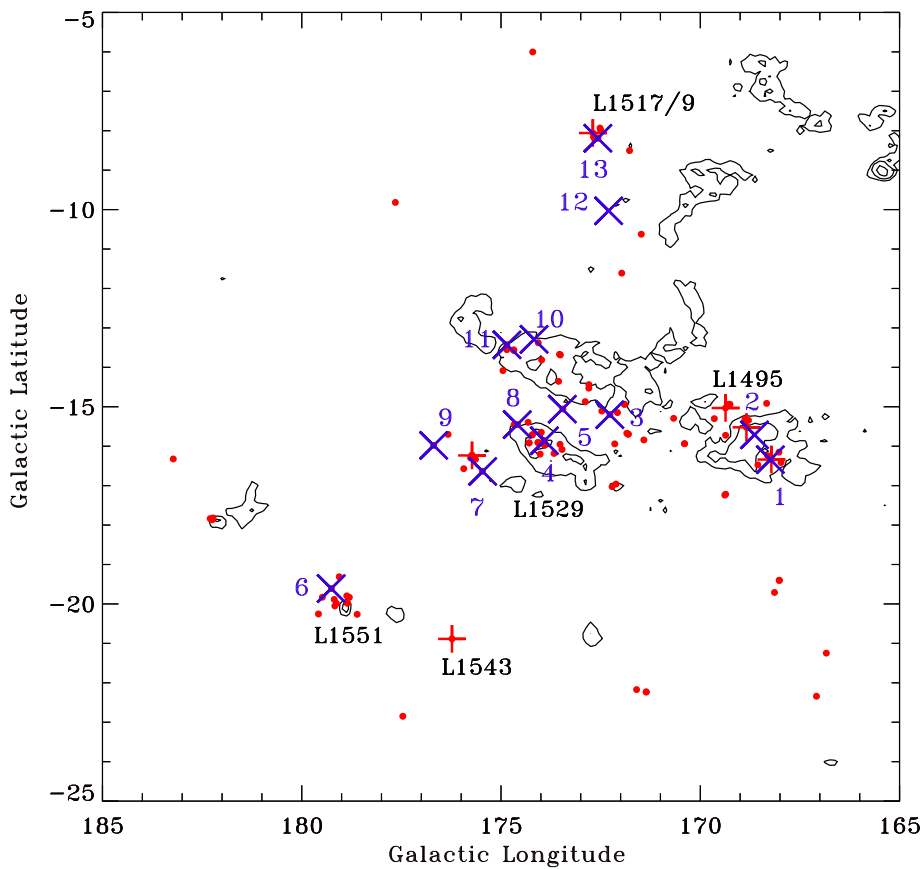} 
\caption{Location of the observed stars in the Taurus Auriga region.
The contours are low lying levels of CO integrated intensity from the
\citet{Dame+etal_2001} survey.
The red dots are stars in common between \citet{Ducourant+etal_2005}
 and \citet{Kenyon+etal_2008} master list of PMS stars in Taurus.
The red crosses show the stars with precise distances (VLBA or VB+SB2)
The blue crosses show the stars with masses measured by disk rotation.
}
\label{fig:location}
\end{figure*}

\section{Discussion: Implications for stellar evolution models}

\begin{table*}
\caption{Dynamical masses for single stars}
\label{tab:singles}
\begin{tabular}{rlllllllllll}
\hline
   & Name    & SpType & T$_\mathrm{eff}$ &  $L_*$ & $M_*$ &  Comments \\
\multicolumn{2}{l}{Number}  &  & (K)       & $\Lsun$       & $\Msun$ & \& References \\
1   &  CW Tau  &  K3 &$4840^{+200}_{-220}$ & $2.42\pm1.40$ & $0.69 \pm 0.14$ & M1, L0,  L1495\\
2   &  CY Tau  & M1  &$3615\pm 65$         & $0.40\pm0.09$ & $0.63\pm 0.05$  &M0, L0,  L1495\\
3   &  IQ Tau  & M0.5&$3765\pm 85$         & $0.81\pm0.26$ & $0.79\pm  0.02$ & M0, L0\\
4   & Haro 6-13& M0  &$3850^{+200}_{-170}$ & $0.69\pm0.22$ & $1.00\pm 0.15$  & M4, L0. L1529\\
5   &  DL Tau  & K7  &$4050^{+150}_{-200}$ & $0.74\pm0.38$ & $0.91 \pm 0.02$ & M0, L0 \\
6   & DM Tau   & M1  &$3680^{+170}_{-130}$ & $0.23\pm0.02$ & $0.53 \pm 0.02$ & M2, L0, L1551\\
7   & CI Tau   & K7  &$4050^{+150}_{-200}$ & $0.93\pm0.35$ & $0.80\pm 0.02$ & M0, L0,  L1529  \\
8   & DN Tau   & M0  &$3850^{+200}_{-170}$ & $0.79\pm0.16$ & $0.95\pm 0.16$ & M0, L0, L1529   \\
9  & LkCa 15   & K5  &$4450^{+170}_{-250}$ & $0.81\pm0.21$ & $1.01\pm 0.03$ & M2, L0 \\
10 & Haro 6-33 & M0.5&$3765^{+115}_{-85}$  & 0.76          & $0.5\pm 0.1$   & M4, L1\\
11 & GO Tau    & M0  &$3850^{+200}_{-170}$ & $0.29\pm0.09$ &$0.48 \pm 0.01$ &M0, L0 \\
12 & DS Tau    & K5  &$4450^{+170}_{-250}$ & $0.76\pm0.34$ &$0.68 \pm 0.12$ &M1, L0 \\
13 & GM Aur    & K3  &$4850^{+200}_{-220}$ & $1.23\pm0.32$ &$1.00 \pm 0.02$ & M3, L0, L1517/9\\
\hline
\end{tabular}
\tablefoot{References for Mass: M0, this work; M1,
\citet{Pietu+etal_2014}; M2, \citet{Pietu+etal_2007};
 M3, \citet{Dutrey+etal_2008}; M4, \citet{Schaefer+etal_2009}.\\
 References for Spectral Type and Luminosity:  L0, \citet{Andrews+etal_2013}; L1, \citet{White+Hillenbrand_2004}
} 
\end{table*}

Comparison of the measured masses  with theoretical
evolutionary tracks is  best made on a modified HRD in which
the distance-independent quantity $L/M_*^2$ is plotted 
$vs~T_\mathrm{eff}$  \citep{Simon+etal_2000}. Table \ref{tab:singles} lists the stellar
parameters of the stars we consider for a comparison with
models of PMS stellar evolution.   For completeness, Table \ref{tab:singles} also 
lists the masses for several of  stars that were reported previously. 
Figure \ref{fig:location} shows the locations of these stars within the SFR.
Of the stars in  Table \ref{tab:vmass}, Table \ref{tab:singles} does not include MWC 480, HV Tau C,  
and 04302+2247. \citet{Pietu+etal_2007} have already carefully 
considered  MWC 480 on the modified HRD.   We will discuss the multiple system 
HV Tau and its masses in a  forthcoming paper (Guilloteau et al 2014, in prep.), 
and 04302+2247 is a Class I YSO with inadequately known stellar parameters
\citep{Graefe+etal_2013}.  All spectral types (col. 3),  effective
temperatures,  $T_\mathrm{eff}$ (col. 4), and luminosities (col. 5)  are from
\citet{Andrews+etal_2013} except for Haro 6-33 which is from \citet{White+Hillenbrand_2004}. 
We assumed   $T_\mathrm{eff}$ uncertainties corresponding 
to $\pm 1$ ~spectral type and used the spectral type-$T_\mathrm{eff}$ look-up 
table in \citet{Pecaut+Mamajek_2013}.  To calculate uncertainties in
$L/M_*^2$ we propagated the uncertainties
in L and M  according to the usual procedure for the uncertainty of
a ratio when one of the variables is squared.

\begin{figure*}
\includegraphics[width=8.5cm]{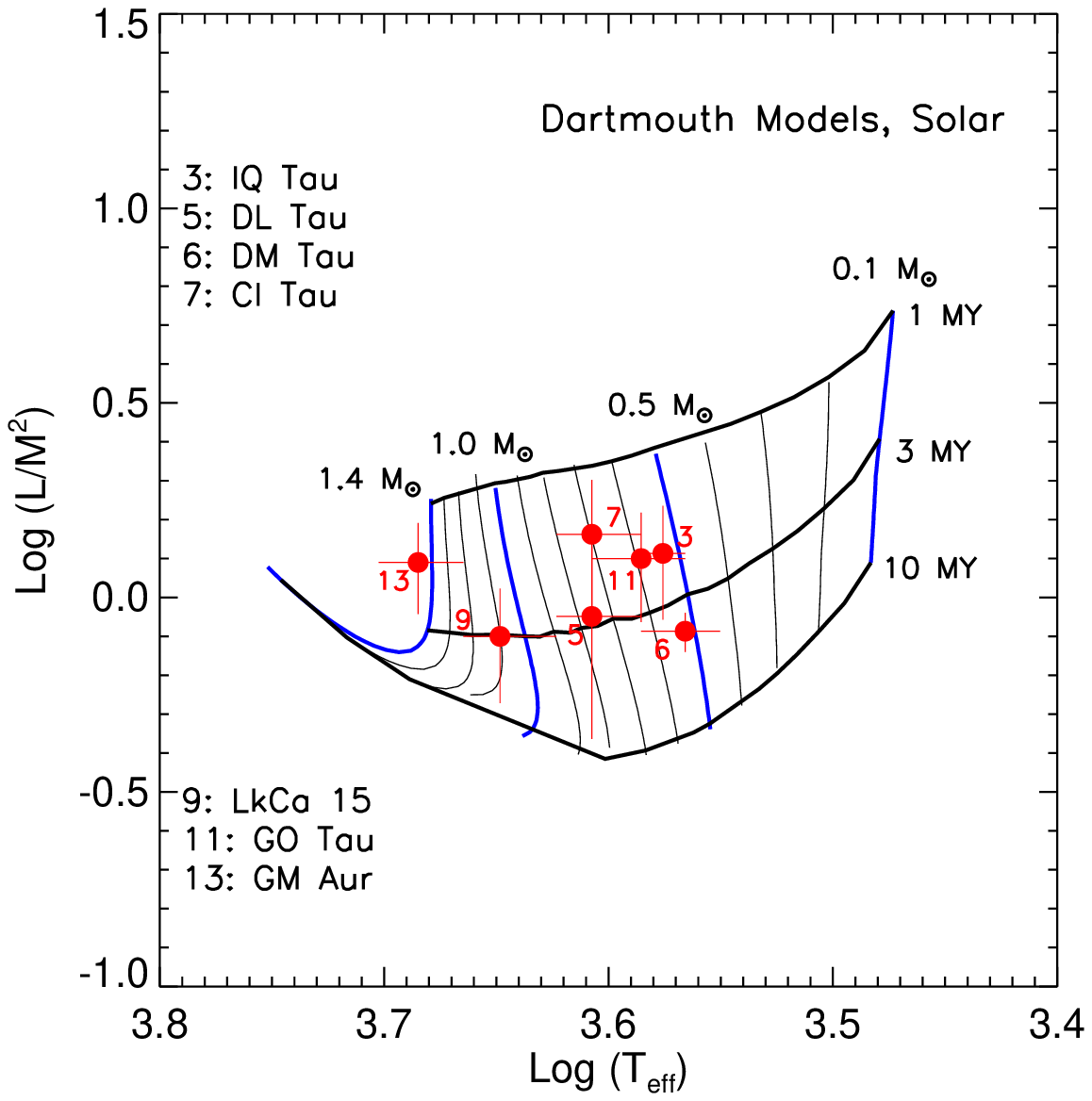}
\includegraphics[width=8.5cm]{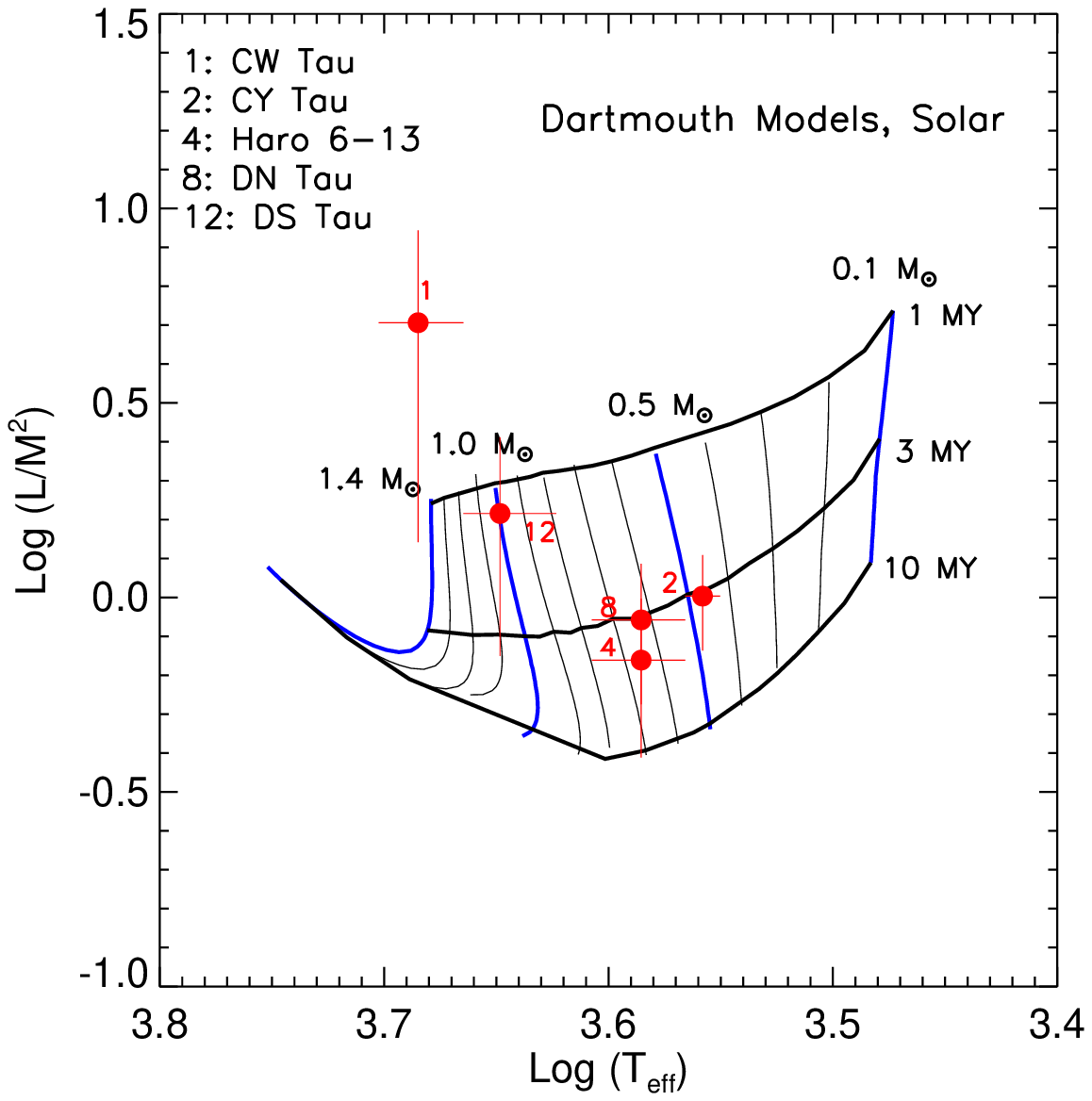}
\caption{Stars on the modified, distance-independent, HR diagram $L/M^2 vs$ T$_\mathrm{eff}$ from
\citet{Dotter+etal_2008} evolutionary tracks. Left: stars with dynamical masses accurate to $<5 \%$,
right: other stars.}
\label{fig:dtmhtracks}
\end{figure*}

\begin{table*} %
\caption{Comparison of best Masses with tracks}
\begin{tabular}{rllllll}
\hline
\#  & Star    & Dynamical     & \multicolumn{4}{c}{Agreement \& Evolutionary Track Mass range} \\
    &        & Mass           &  BCAH      & SDF       & Pisa      & Dartmouth \\
~3. & IQ Tau  &$ 0.79\pm0.02$  & F~~0.65-0.70  & F~~0.45-0.70 & F~~0.48-0.72 & P~~0.50-0.60 \\       
~5. & DL Tau &$ 0.91\pm0.02$   & E~~0.72-0.92  & E~~0.60-0.92 & F~~0.62-0.88 & F~~0.68-0.88\\
~6. & DM Tau &$ 0.53\pm0.02$   &  E~~0.50-0.70 & E~~0.35-0.59 & E~~0.43-0.62 & E~~0.44-0.62\\
~7. & CI Tau & $0.80\pm0.02$   &  E~~0.70-0.88 & E~~0.58-0.90 & E~~0.60-0.82 & E~~0.59-0.80\\
~9. & LkCa 15&$1.01\pm0.03$    &  E~~0.92-1.30 & E~~0.95-1.30 & E~~0.90-1.25 & E~~0.92-1.30\\
11. & GO Tau &$0.48\pm0.01$    & P~~0.70-0.92  & E~~0.45-0.75 & F~~0.52-0.82 & E~~0.48-0.70\\
13. & GM Aur &$1.00\pm0.02$    &   P~~1.35-?   &  P~~1.51-?   &  P~~1.30-?   &  P~~1.20-?\\
\hline
\end{tabular}
\label{tab:tracks} 
\tablefoot{Qualitative agreement:
E= Excellent G= Good F= Fair P= Poor}
\end{table*}

For stars in Table \ref{tab:singles} with mass uncertainties  less than or equal to
 $\pm 5\%$, Fig.\ref{fig:dtmhtracks}a plots their  $L/M_*^2$ {\it vs} $T_\mathrm{eff}$  on a 
modified HRD  using the  evolutionary tracks calculated by \citet[][Dartmouth]{Dotter+etal_2008}.
Figure \ref{fig:dtmhtracks}b is a similar plot for stars with mass precisions
worse than $\pm 5\%$.  Figure \ref{fig:dtmhtracks}b does not include Haro 6-33 because
the uncertainty of its luminosity is not available.   Figs.\ref{fig:bcahtracks}-\ref{fig:pisatracks}
in Appendix \ref{app:tracks} show similar plots for the evolutionary
tracks of \citet[][BCAH]{Baraffe+etal_1998}, \citet[][SDF]{Siess+etal_2000}, 
and \citet[][Pisa]{Tognelli+etal_2011}. The numbering of the stars follows that of Table \ref{tab:singles}.

Three features are apparent in all the modified HRD plots:
\begin{enumerate}\itemsep 0pt
\item
 No stars with mass below $\sim 0.5\Msun$ appear.
Mass measurement by the circumstellar disk technique of lower mass 
stars in Taurus requires higher sensitivity, presumably because their 
disks are less massive \citep[e.g.][]{Schaefer+etal_2009,Andrews+etal_2013} 
but also much smaller \citep{Pietu+etal_2014}.
\item
$T_\mathrm{eff}$ uncertainties severely compromise the comparisons 
with the tracks because most uncertainties span tracks separated by more 
than $0.1 \Msun$. 
\item
Uncertainties in  $L/M_*^2$  affect comparisons with  evolutionary
tracks  very little because, at the ages  displayed, the tracks are nearly 
parallel to the   $L/M_*^2$ axis. However, uncertainties in  $L/M_*^2$ can 
affect the age estimate.
\end{enumerate}

Table \ref{tab:tracks} summarizes, qualitatively, a comparison of the best masses with the 4 
theoretical tracks.  Columns 1 and 2 identify the stars and column 3
repeats, for convenience, their measured dynamical masses at 140 pc
distance.  Columns 4-7 give the
mass range corresponding to the $\pm 1 \sigma$ uncertainties of the ($T_{eff}, L/M_*^2$)
values.   Table \ref{tab:tracks} indicates that the measured masses of 
DM Tau, CI Tau, and LkCa 15 (0.53, 0.80, and $1.01 \Msun$, respectively) are 
in agreement with all 4 sets of tracks if they are actually at the fiducial distance
140\,pc.   
However, IQ Tau with a measured mass indistinguishable from that
of CI Tau, and GM Aur with a measured mass indistinguishable from that
of LkCa 15, are inconsistent with their nominal mass tracks $0.80 \Msun$ and
 $1.00 \Msun$, respectively.
 Two reasons are likely.  Their published
spectral types may not provide an accurate indication of their $T_\mathrm{eff}$.
Also, their distances may be greater or less than 140 pc which would  
not affect their  $L/M_*^2$ values but would change their mass derived
from the angular measure of their disk rotation.  For example, Fig.\ref{fig:location} 
shows that the position of GM Aur in the L1517/19 region  
is close to NTTS 045251+3016 at distance $158.7\pm3.9$~pc \citep{Simon+etal_2013}.
If this distance applies to GM Aur, its mass would become (158.7/140)
times greater, i.e. $1.13\pm 0.02 \Msun$, decreasing the discrepancy
with respect to the $1.1 \Msun$ theoretical track but still more than
$\sim 1 \sigma$ too high with respect to all the $1.1 \Msun$ tracks.
Fig.\ref{fig:location} also shows that the position of CI Tau in the L1529 region  
is close to HP Tau G2 for which \citet{Torres+etal_2009} measured the precise 
distance $161.2\pm 0.9$ pc. At this distance the measured mass of CI Tau
would be (161.2/140) times greater or $0.92\pm 0.02 \Msun$.  If this were its
actual mass, the quality of consistency with the Dartmouth and Pisa evolutionary
would be degraded. It seems reasonable to conclude that a definitive comparison
of the measured masses with the available theoretical evolutionary 
tracks should await a more accurate determination of effective temperatures of
the target stars, mass measurements of stars with masses below  $0.5\Msun$,
and accurate distances to all.

For the stars with the most precise mass measurements, Table \ref{tab:ages} summarizes 
the ages indicated by their locations on the modified HRDs.  All 4 theoretical
calculations indicate the stars are younger than 10 Myr, and that a representative
age for most of the stars is between 2 and 4 Myr, consistent with prior estimates
\citep[e.g.][]{Kenyon+etal_2008}.  Improved values of stellar parameters will yield
more accurate age estimates.

In conclusion, the reliability of these comparisons of the measured masses with the
theoretical  tracks is compromised by the  uncertainties on the  effective
temperatures and distances.  Also, a definitive assessment of the
theoretical evolutionary tracks must await mass measurements of 
stars with masses below $0.5 \Msun$.   We look ahead to {\it GAIA} to provide
precise distances to stars in the Taurus star forming region and
have started measurements that we hope will improve the
determinations of $T_\mathrm{eff}$ and will extend mass measurements to
masses smaller than those presented in this paper.

\begin{table}
\caption{Approximate ages (Myr)}
\begin{tabular}{rllllll}
\hline
\# & Star    & \multicolumn{4}{c}{Evolutionary Tracks}\\
   &        &  BCAH & SDF & Pisa & Dartmouth \\
\hline
3.  & IQ Tau  &  2 &    3 &      3 &      2 \\
5.  & DL Tau  &  3 &    4 &      4 &      3  \\
6.  & DM Tau  &  5 &    8 &      5 &     3.5  \\
7.  & CI Tau  & 1.5 & 2.5 &    2.5 &    1.5     \\
9.  & LkCa 15 &  3 &    4 &      4 &      3      \\
11. & GO Tau  &  2 &    4 &      3 &      3     \\
13. & GM Aur  &  2 &    2 &      2 &    1.5    \\
\hline
\end{tabular}
\label{tab:ages} 
\end{table}

\section{Conclusion}

We have analyzed $\sim 1.2''$ arcsecond resolution observations of the CN N=2-1
line in a sample of T Tauri stars, and performed a comparison with
$\sim 0.6''$ resolution data obtained in $^{12}$CO for several sources.
The results show that the CN transition is a good, sensitive, contamination free, dynamical mass tracer for stars
in the M1 - A4 spectral type range that are surrounded by disks larger than about 250 AU. The striking ability
of CN to overcome the contamination  problem is best shown by the detection of the disk
in IQ Tau that escaped detection in CO at similar angular resolution.

The largest uncertainty in the mass derivation comes from
the determination of the inclination, although the impact is small for sources with $i > 45^\circ$.
This uncertainty can be reduced by higher angular observations, at the expense
of more observing time.
    Another pending problem is that the more embedded, presumably younger, objects like DG Tau-B
    or IRAS04302+2247, have more complex structure and CN does not appear to be
    a very good disk tracer in these sources.

We used the derived dynamical masses in conjunction with similar
measurements performed using other tracers to compare with the evolutionary tracks.
Agreement of the measured masses with the evolutionary tracks is quite good.  The
                    discrepancies are most likely attributable to inaccurate
                    effective temperatures and, to a lesser extent, distances different  from
                    the 140 pc average value.  We hope to remove some
                    of these discrepancies and to extend the mass
                    measurements below $0.5 \Msun$ (spectral type earlier than M2) where the differences 
                    among the theoretical calculations are the
                    greatest.


\begin{acknowledgements}
This work was supported by ``Programme National de Physique Stellaire'' (PNPS) and ``Programme
National de Physique Chimie du Milieu Interstellaire'' (PCMI) from INSU/CNRS. The work of MS
was supported in part by NSF grant AST 09-07745.
This research has made use of the SIMBAD database,
operated at CDS, Strasbourg, France.
\end{acknowledgements}

\bibliographystyle{aa}

\bibliography{mays-cn}

\appendix
\onecolumn

\section{CN images and profiles}
\label{app:figures}

\begin{figure*}[!h] 
\includegraphics[width=18.0cm]{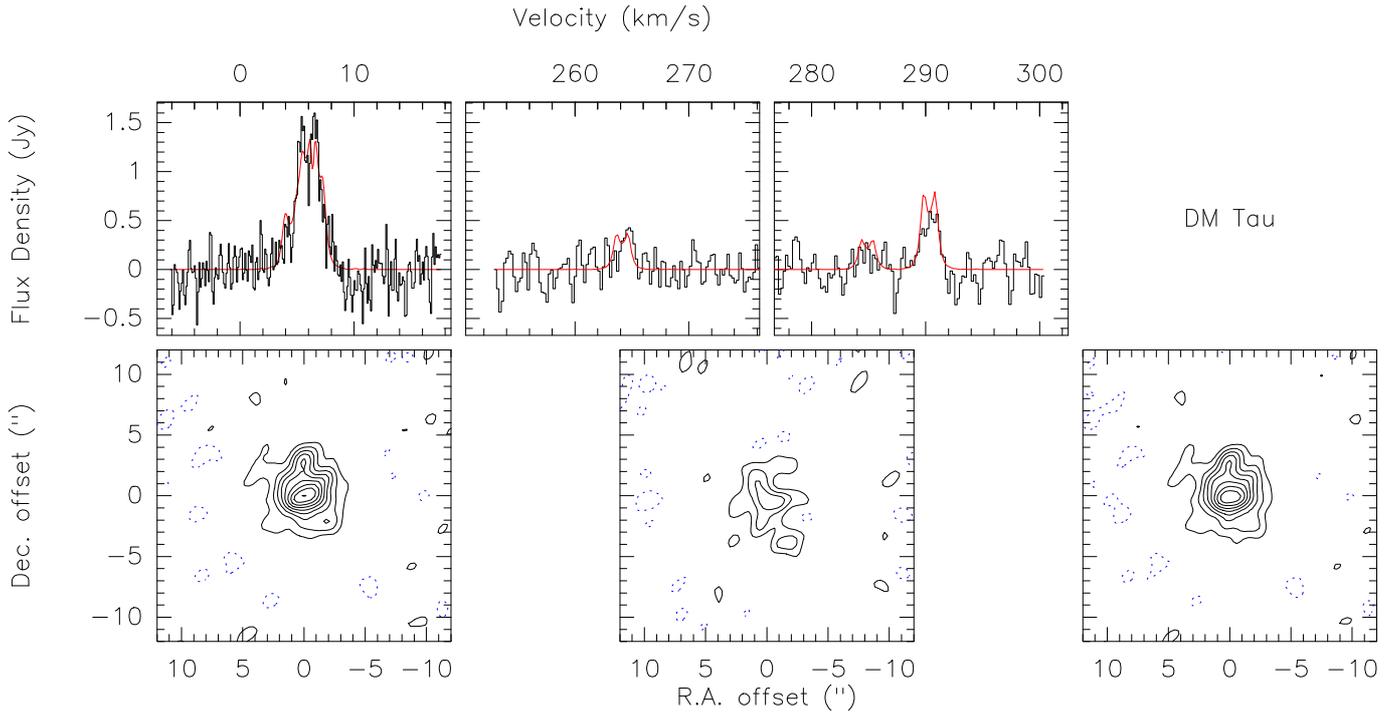}
\caption{Results for DM Tau. Top: Integrated line flux for CN N=2-1 hyperfine components
(histogram) with the best fit profile superimposed (red curve). The spectral resolution
for the strongest group of hyperfine components is 0.103 km\,s$^{-1}$.
Bottom: Signal to noise maps for the two groups of components, and
for all observed components (rightmost panel). Contours are in steps
of 2 $\sigma$.
}
\label{fig:dm_tau}
\end{figure*}

\begin{figure*} 
\includegraphics[width=18.0cm]{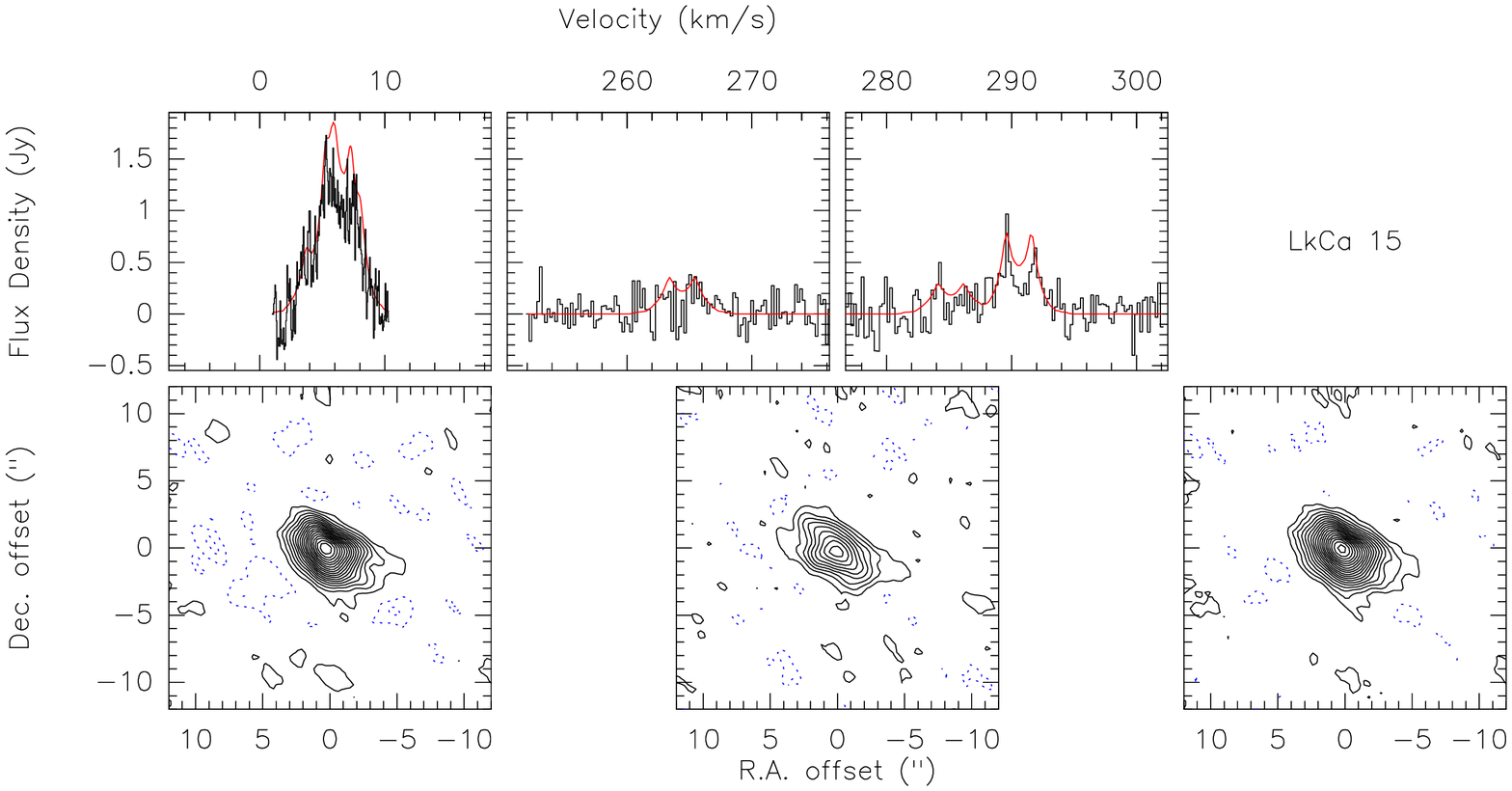}
\caption{As Fig.\ref{fig:dm_tau} for LkCa 15.}
\label{fig:lkca_15}
\end{figure*}

\begin{figure*} 
\includegraphics[width=18.0cm]{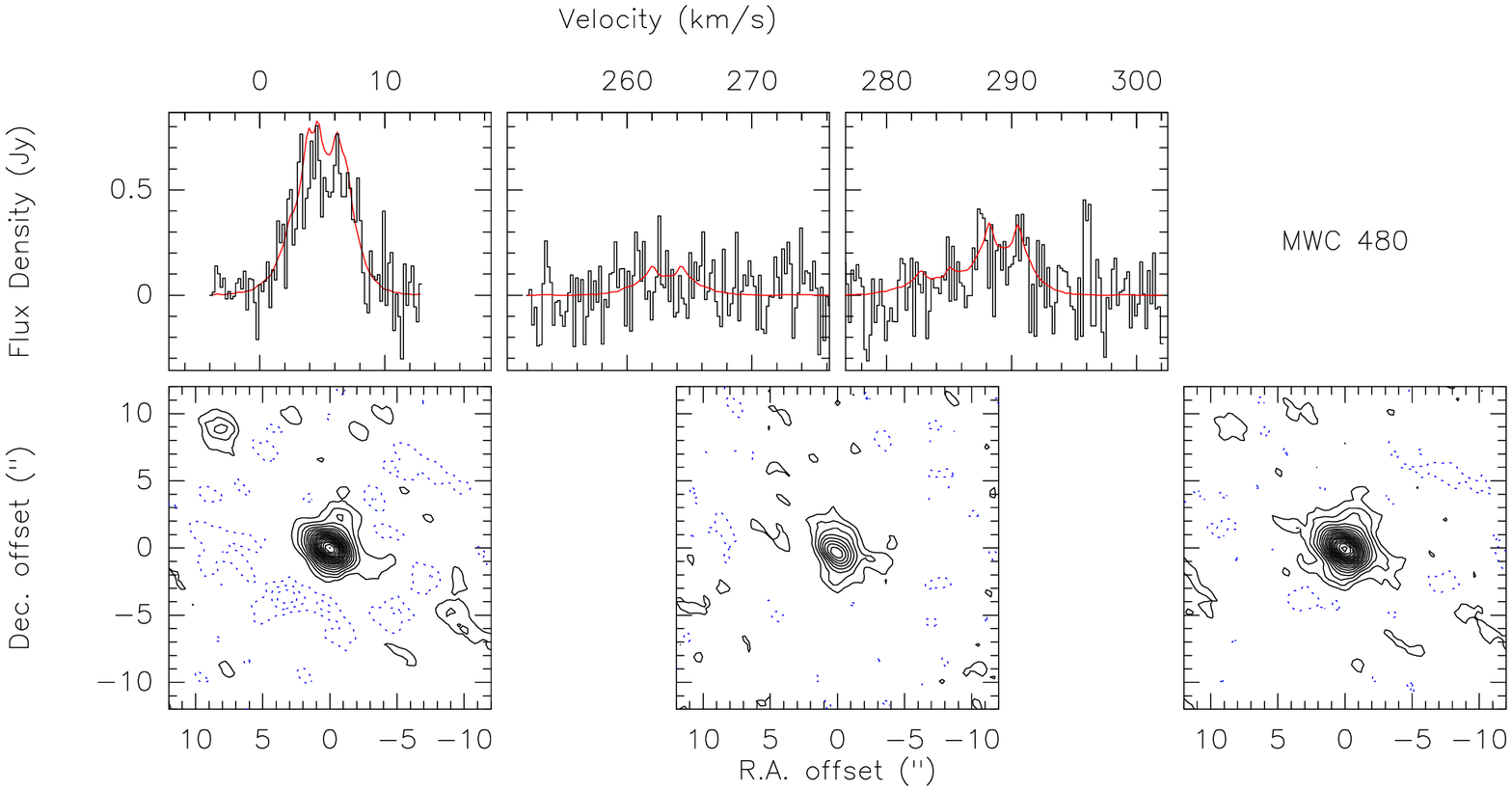}
\caption{As Fig.\ref{fig:dm_tau} for MWC 480.}
\label{fig:mwc_480}
\end{figure*}

\begin{figure*} 
\includegraphics[width=18.0cm]{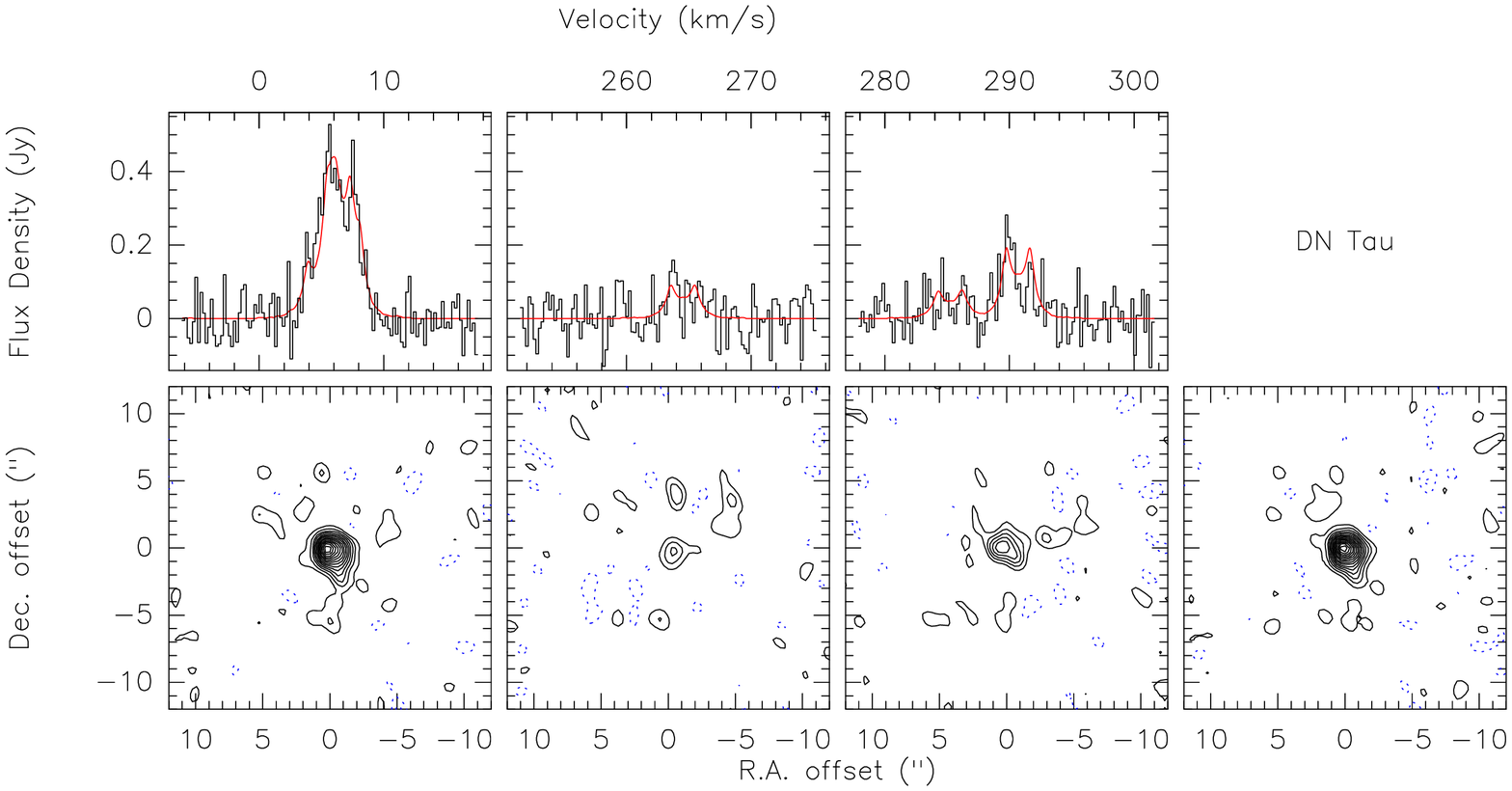}
\caption{Results for DN Tau. Top: Integrated line flux for CN N=2-1 hyperfine components
(histogram) with the best fit profile superimposed (red curve). 
Bottom: Signal to noise maps for each group of components, and
for all observed components (rightmost panel). Contours are in steps
of 2 $\sigma$.
}
\label{fig:dn_tau}
\end{figure*}

\clearpage
\begin{figure*} 
\includegraphics[width=18.0cm]{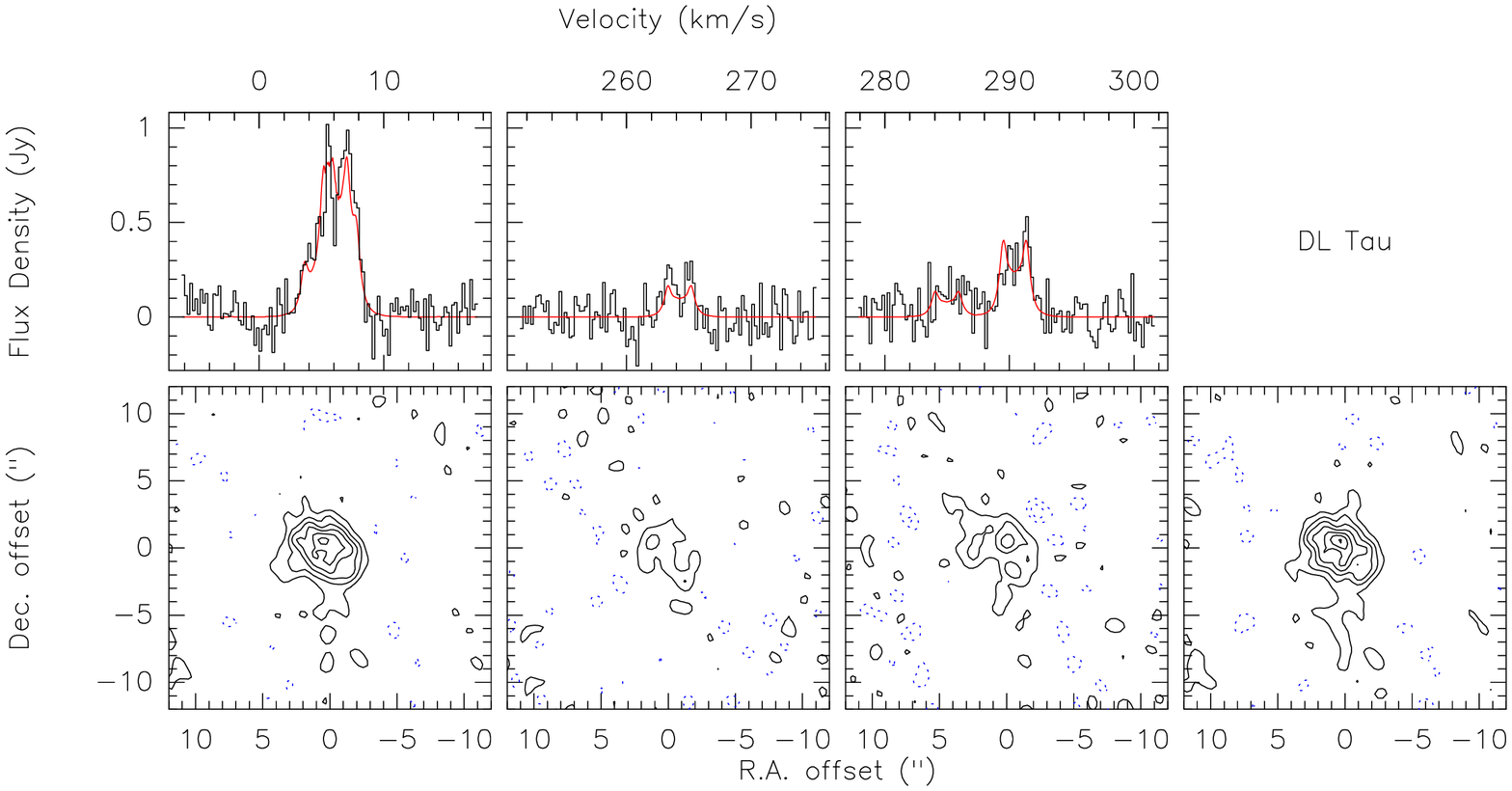}
\caption{As Fig.\ref{fig:dn_tau} for DL Tau.}
\label{fig:dl_tau}
\end{figure*}

\begin{figure*} 
\includegraphics[width=18.0cm]{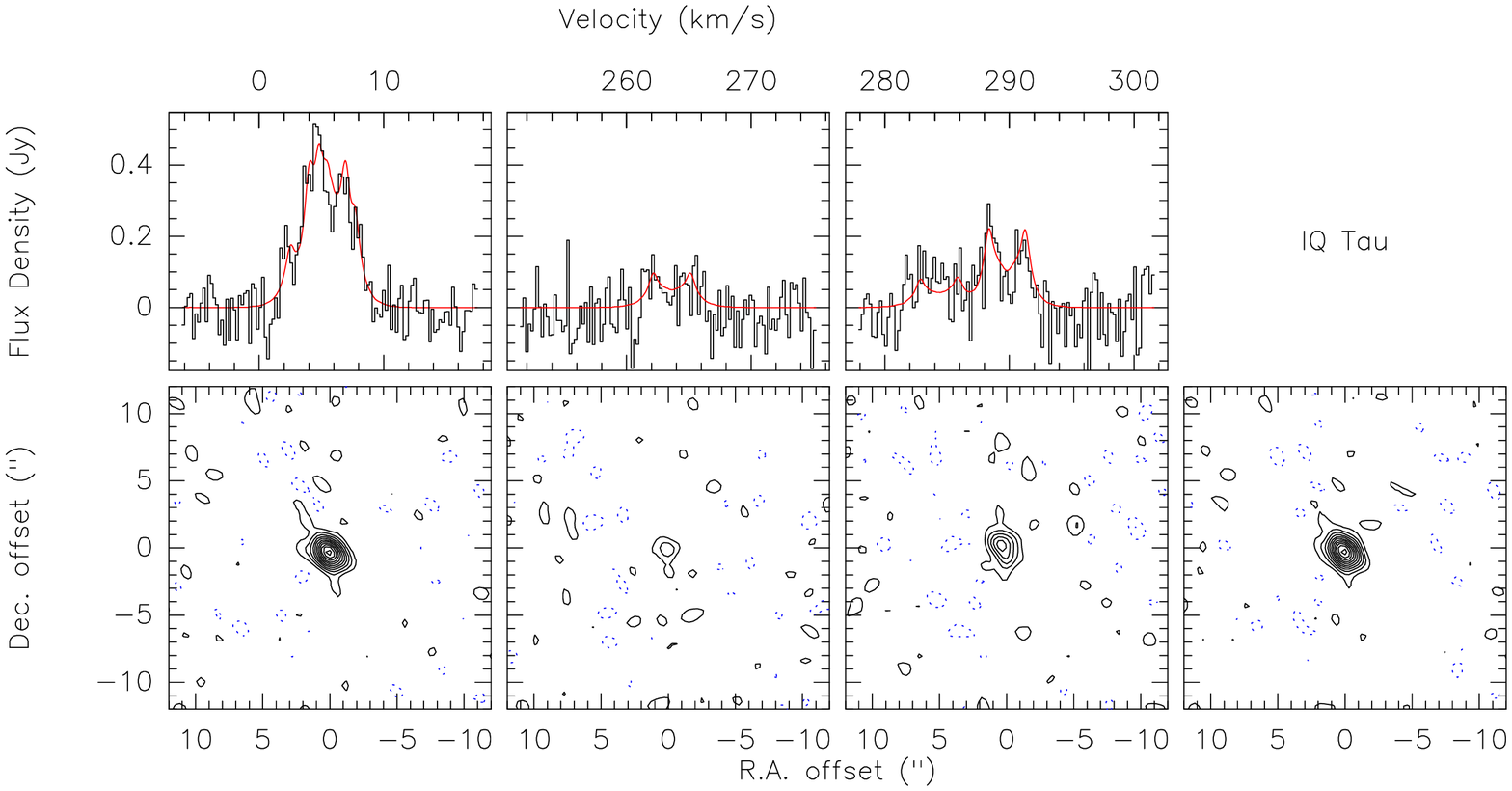}
\caption{As Fig.\ref{fig:dn_tau} for IQ Tau.}
\label{fig:iq_tau}
\end{figure*}

\begin{figure*} 
\includegraphics[width=18.0cm]{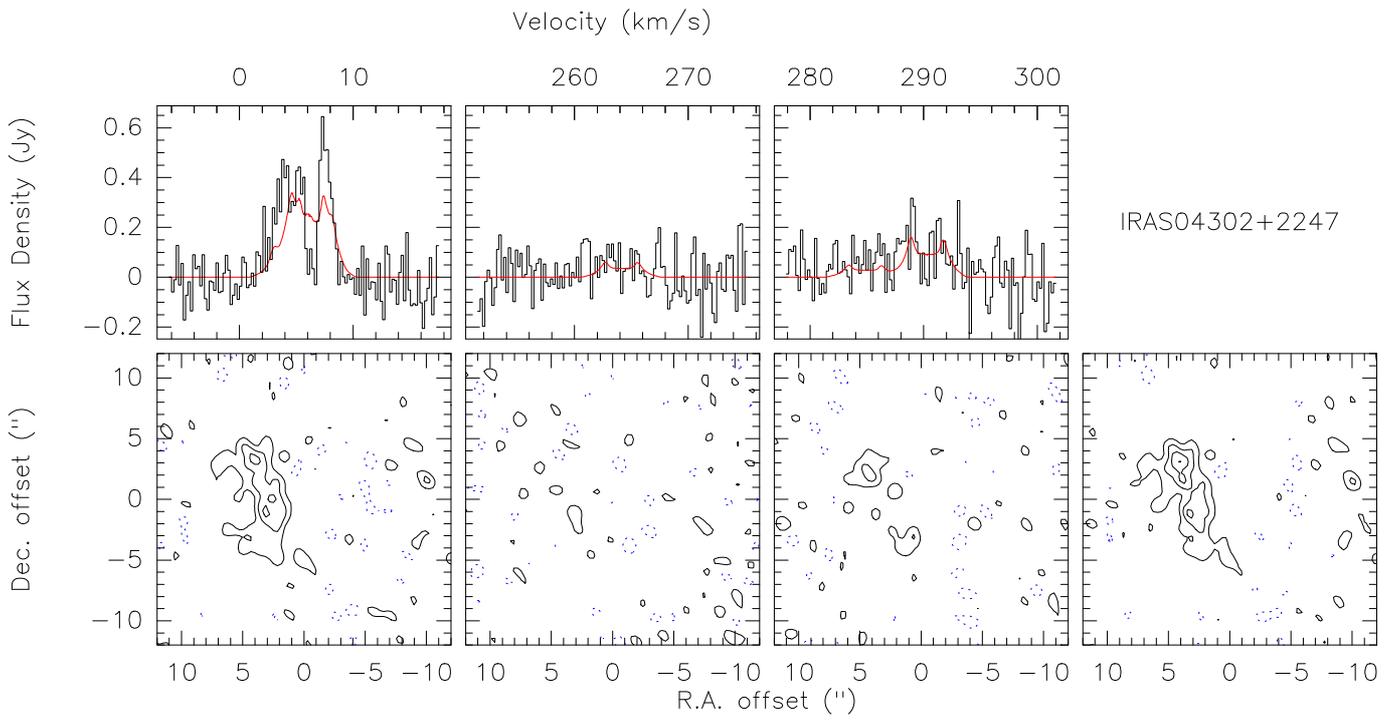}
\caption{As Fig.\ref{fig:dn_tau} for IRAS 04302+2247.}
\label{fig:butter}
\end{figure*}

\begin{figure*} 
\includegraphics[width=18.0cm]{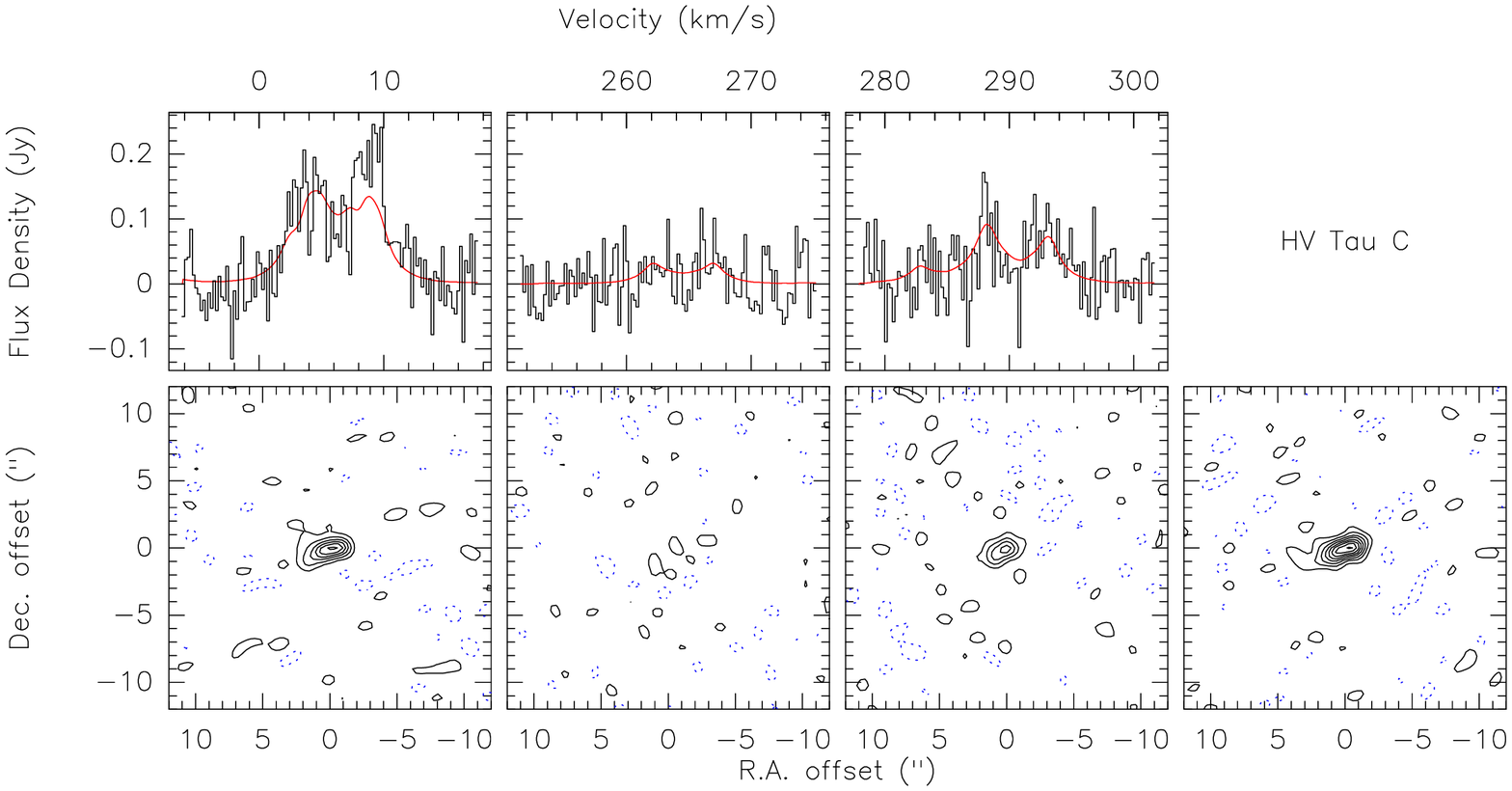}
\caption{As Fig.\ref{fig:dn_tau} for HV Tau.}
\label{fig:hv_tau}
\end{figure*}

\begin{figure*} 
\includegraphics[width=18.0cm]{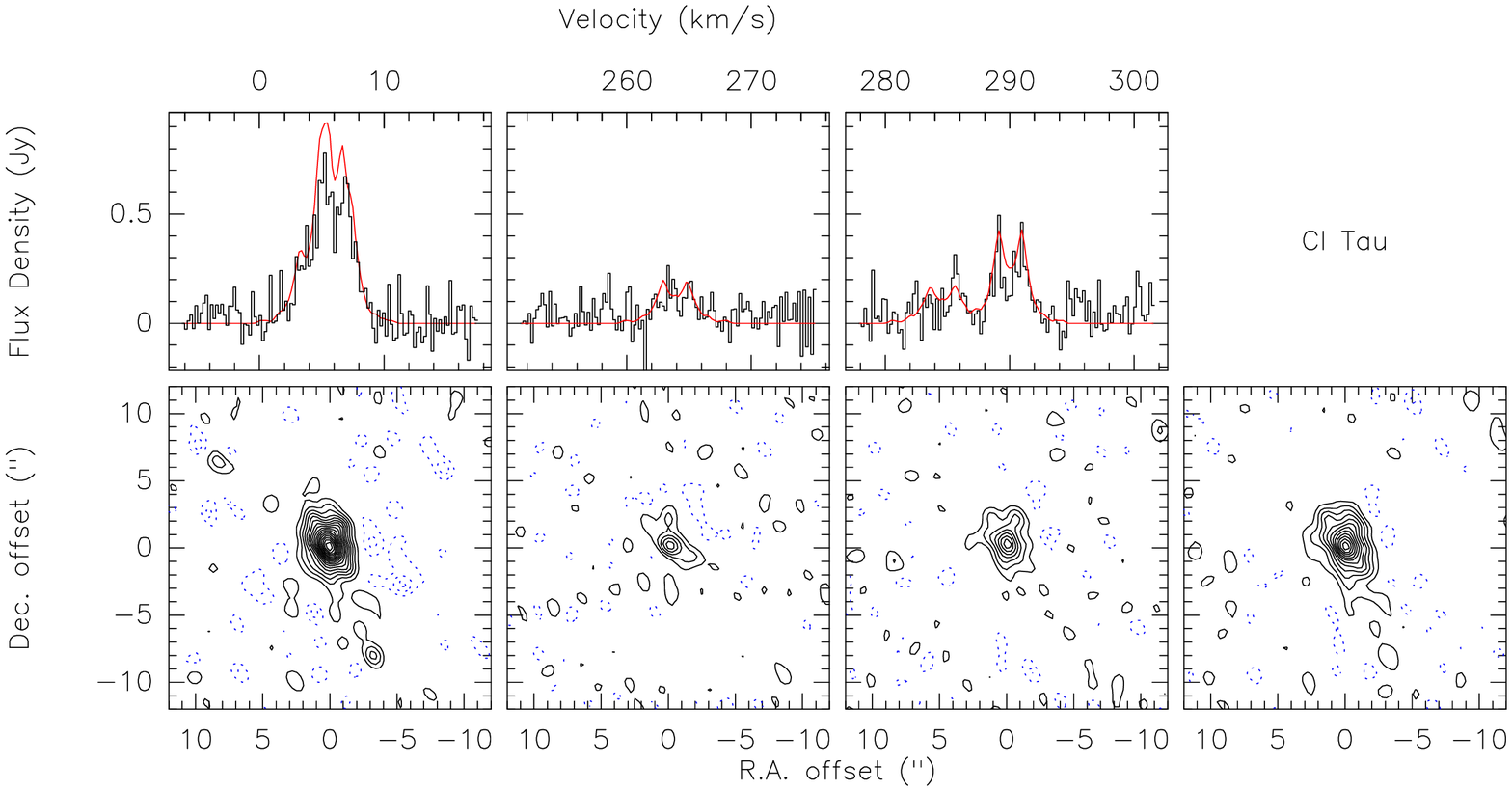}
\caption{As Fig.\ref{fig:dn_tau} for CI Tau.}
\label{fig:ci_tau}
\end{figure*}

\begin{figure*} 
\includegraphics[width=18.0cm]{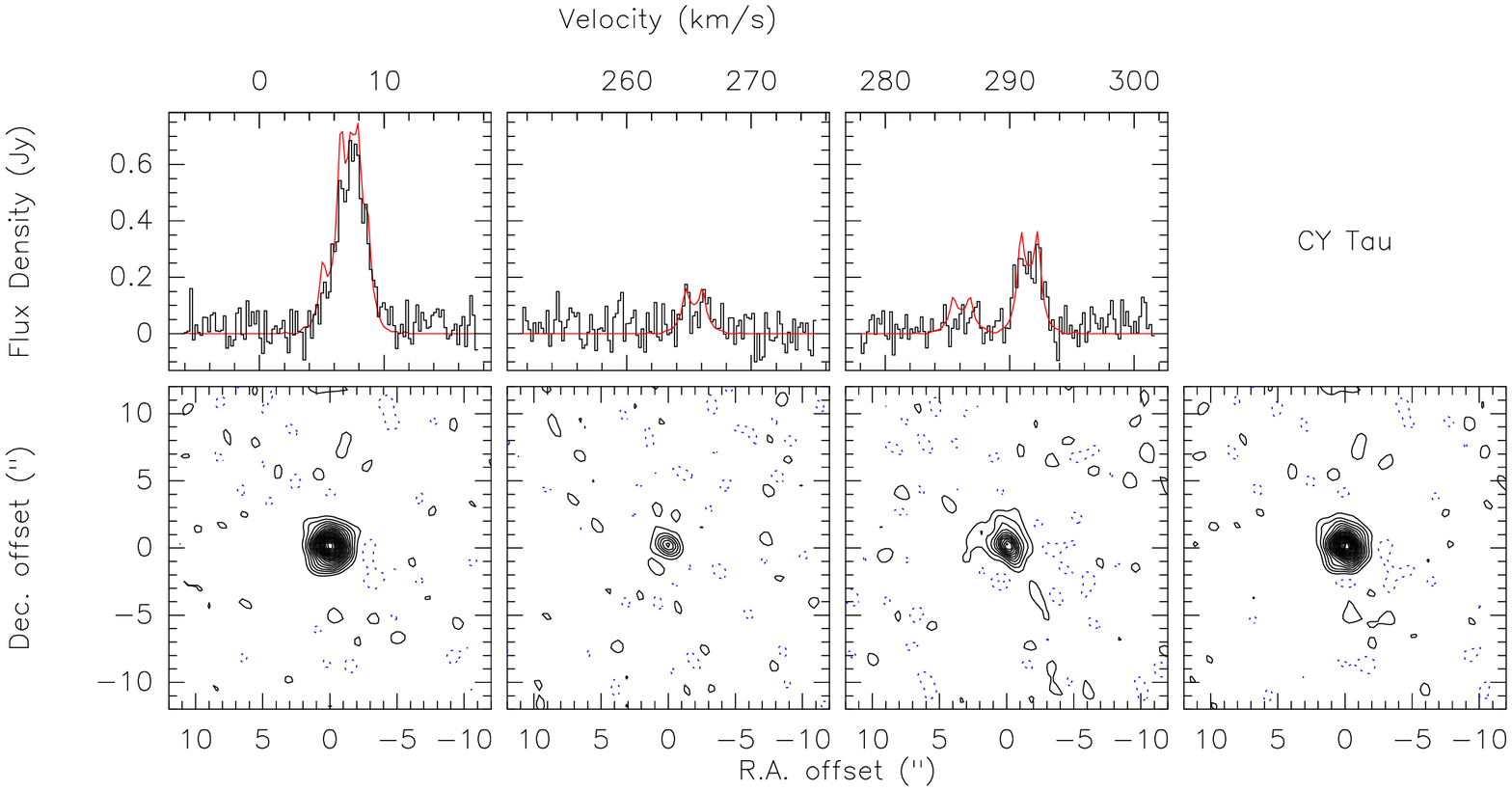}
\caption{As Fig.\ref{fig:dn_tau} for CY Tau.}
\label{fig:cy_tau}
\end{figure*}

\begin{figure*} 
\includegraphics[width=18.0cm]{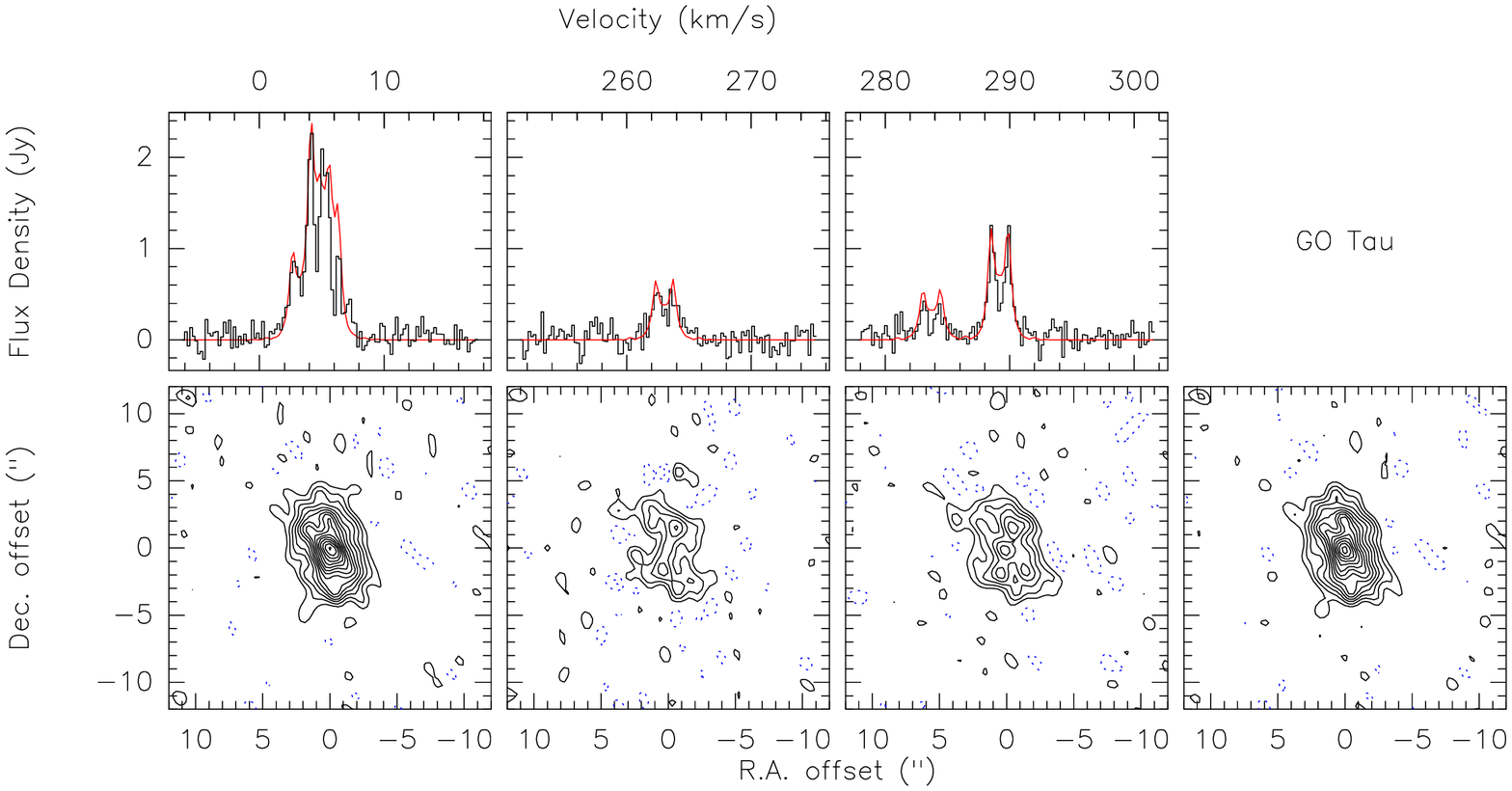}
\caption{As Fig.\ref{fig:dn_tau} for GO Tau.}
\label{fig:go_tau}
\end{figure*}

\clearpage
\section{Fit residuals}
\begin{figure*}[!h] 
\includegraphics[width=18.0cm]{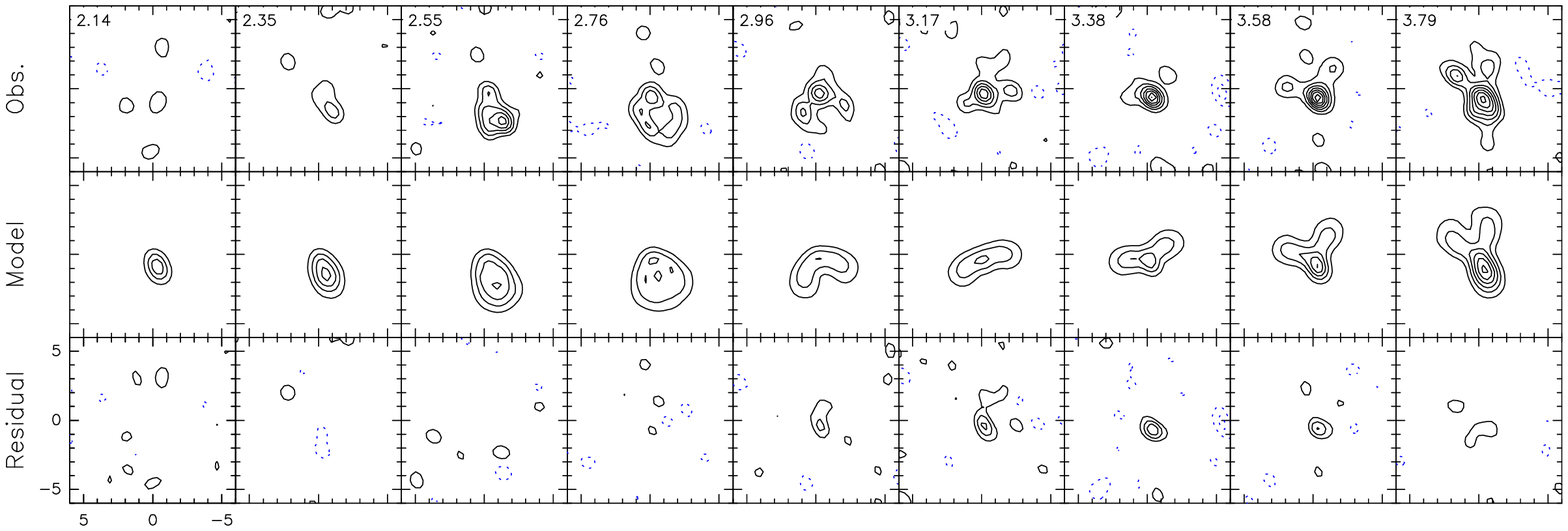}
\includegraphics[width=18.0cm]{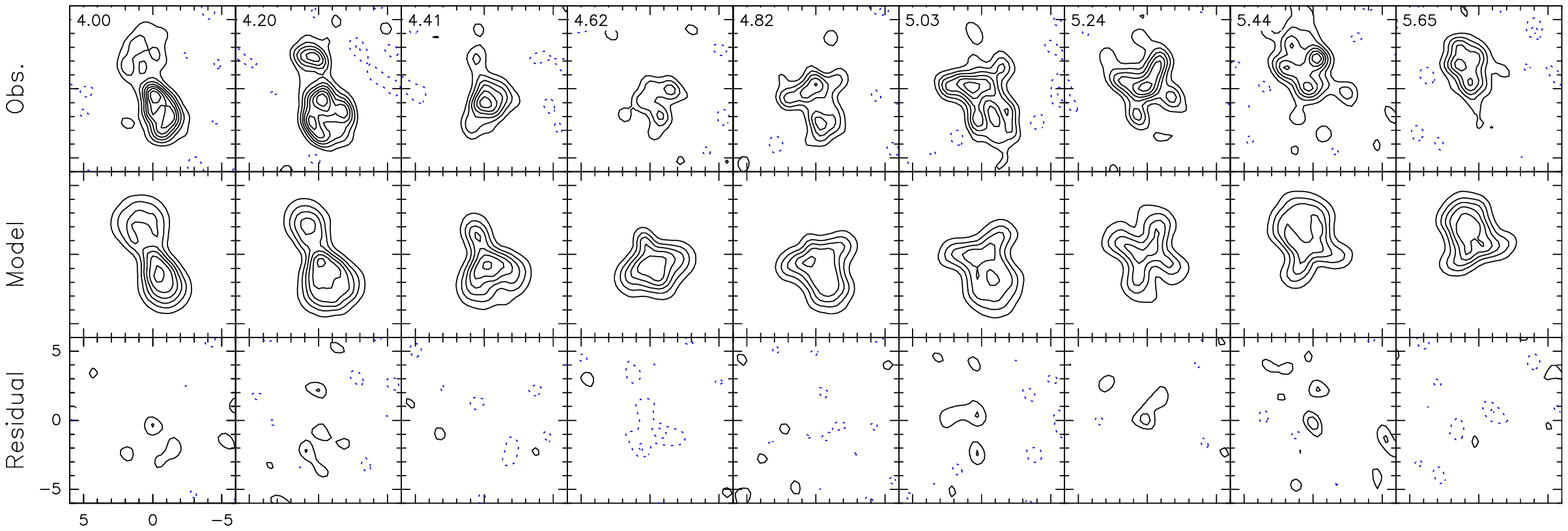}
\includegraphics[width=18.0cm]{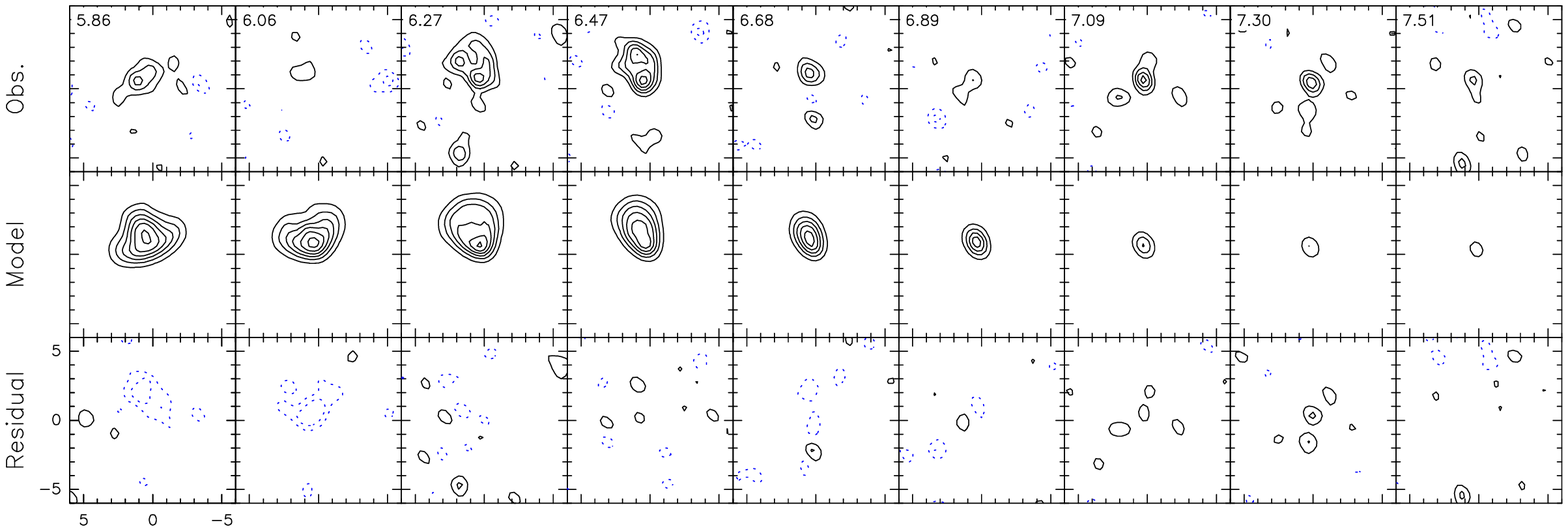}
\caption{Line channels for GO Tau. For each panel, observations
are displayed in the top row, the best fit model in the middle row
and the residuals in the bottom row.  Channels are ordered by
increasing velocities (see upper left corner of each panel)
from left to right and top to bottom.
Contour spacing is 40 mJy/beam, or 2 $\sigma$.  The unusual
spatial distribution is due to the three hyperfine components. The
emission is essentially that of 3 Keplerian disks 
with systemic velocities 3.37, 4.87, and 5.66 
km.s$^{-1}$  and relative intensities 1, 2.67 and 1.68
respectively.}
\label{fig:quality}
\end{figure*}

\newpage
\clearpage

\section{Evolutionary Tracks}
\label{app:tracks}

\begin{figure*}[h]
\includegraphics[width=8.5cm]{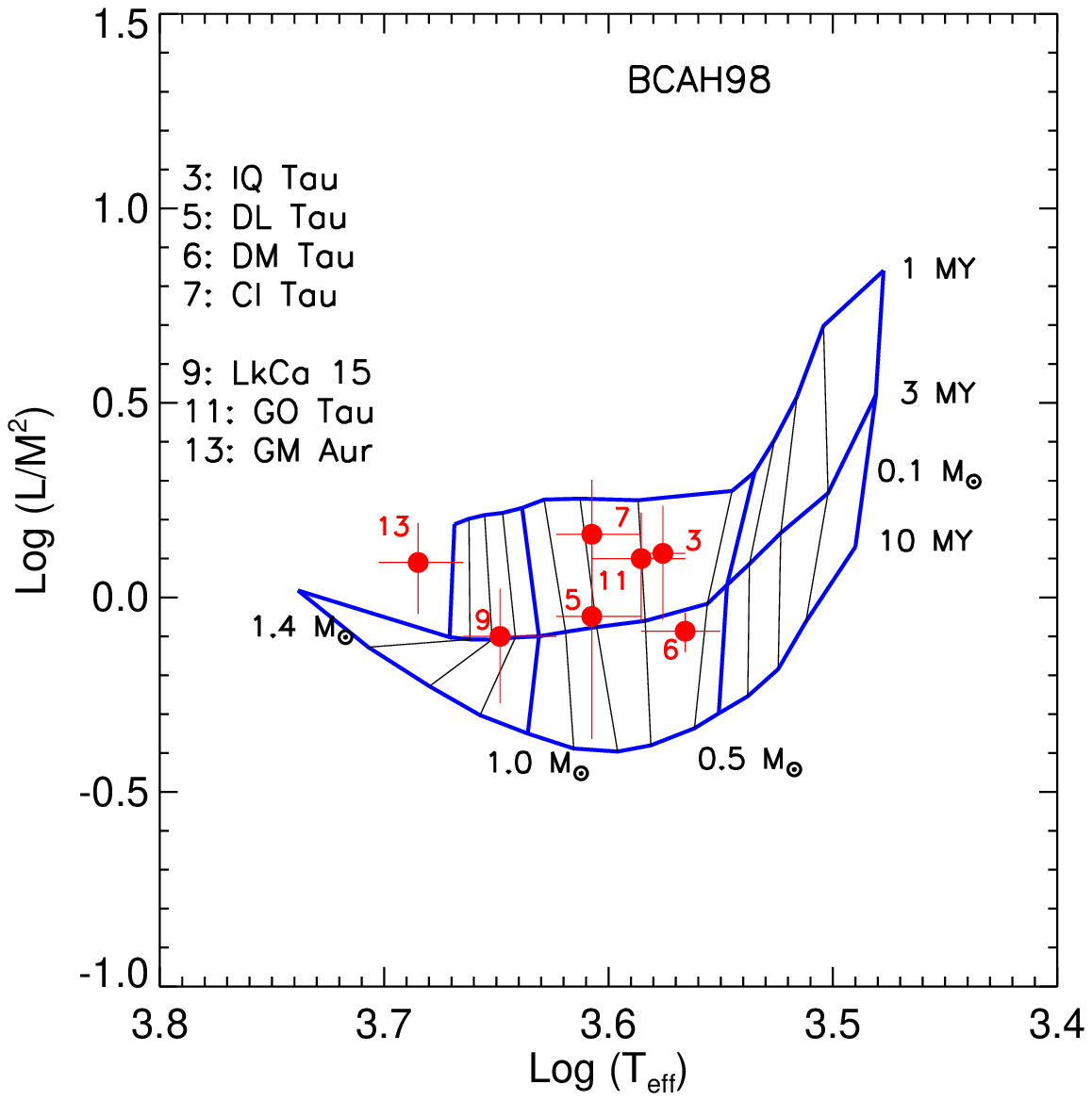}
\hspace{0.50cm}
\includegraphics[width=8.5cm]{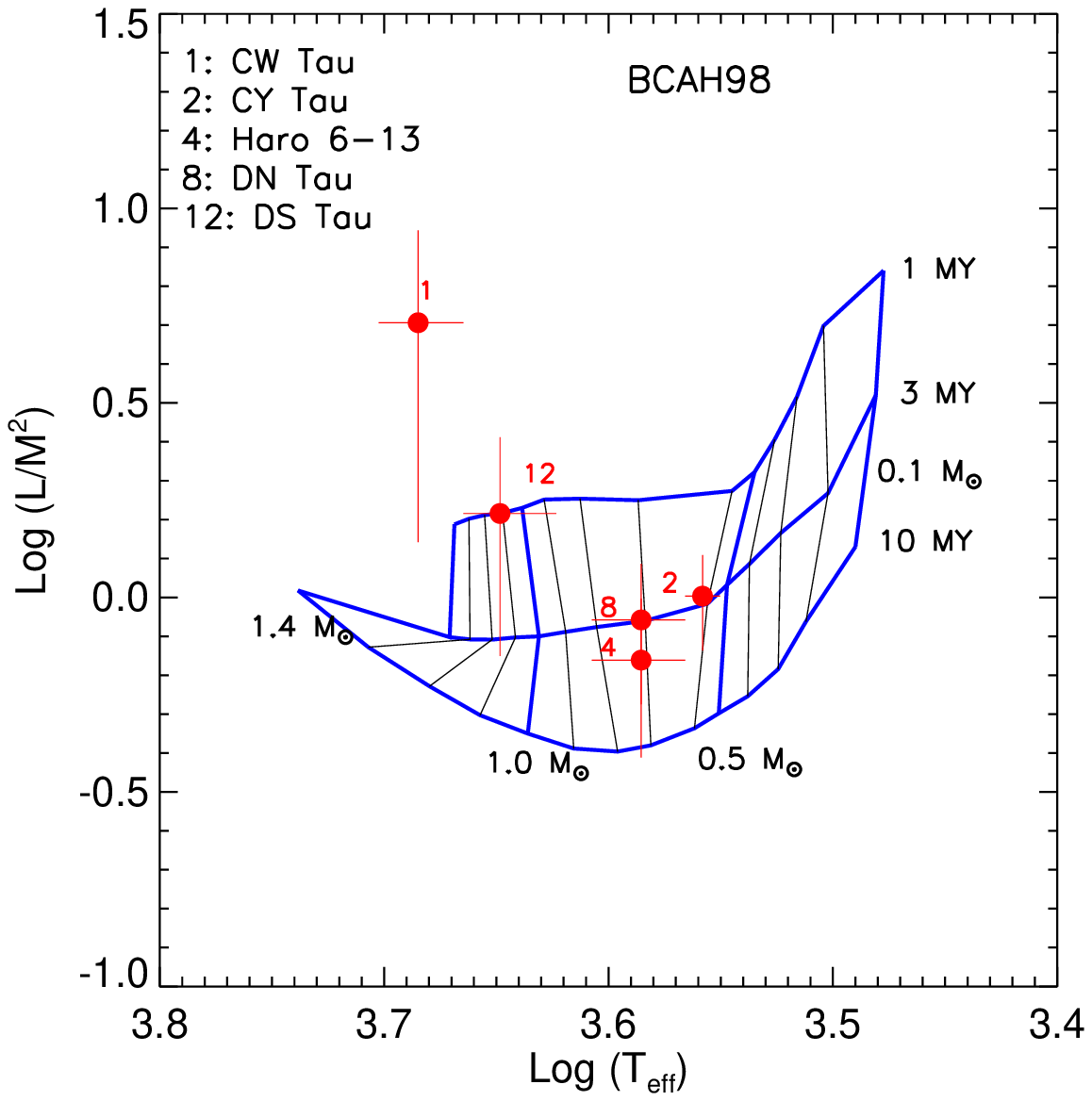}
\caption{Stars on the modified, distance-independent, HR diagram $L/M^2 vs$ T$_\mathrm{eff}$ 
for the \citet{Baraffe+etal_1998} tracks. Stars with dynamical mass precisions
 measured to better than 5\% appear in the left panel
and stars with lower precions appear in the right panel.
}
\label{fig:bcahtracks}
\end{figure*}

\begin{figure*}[h]
\includegraphics[width=8.5cm]{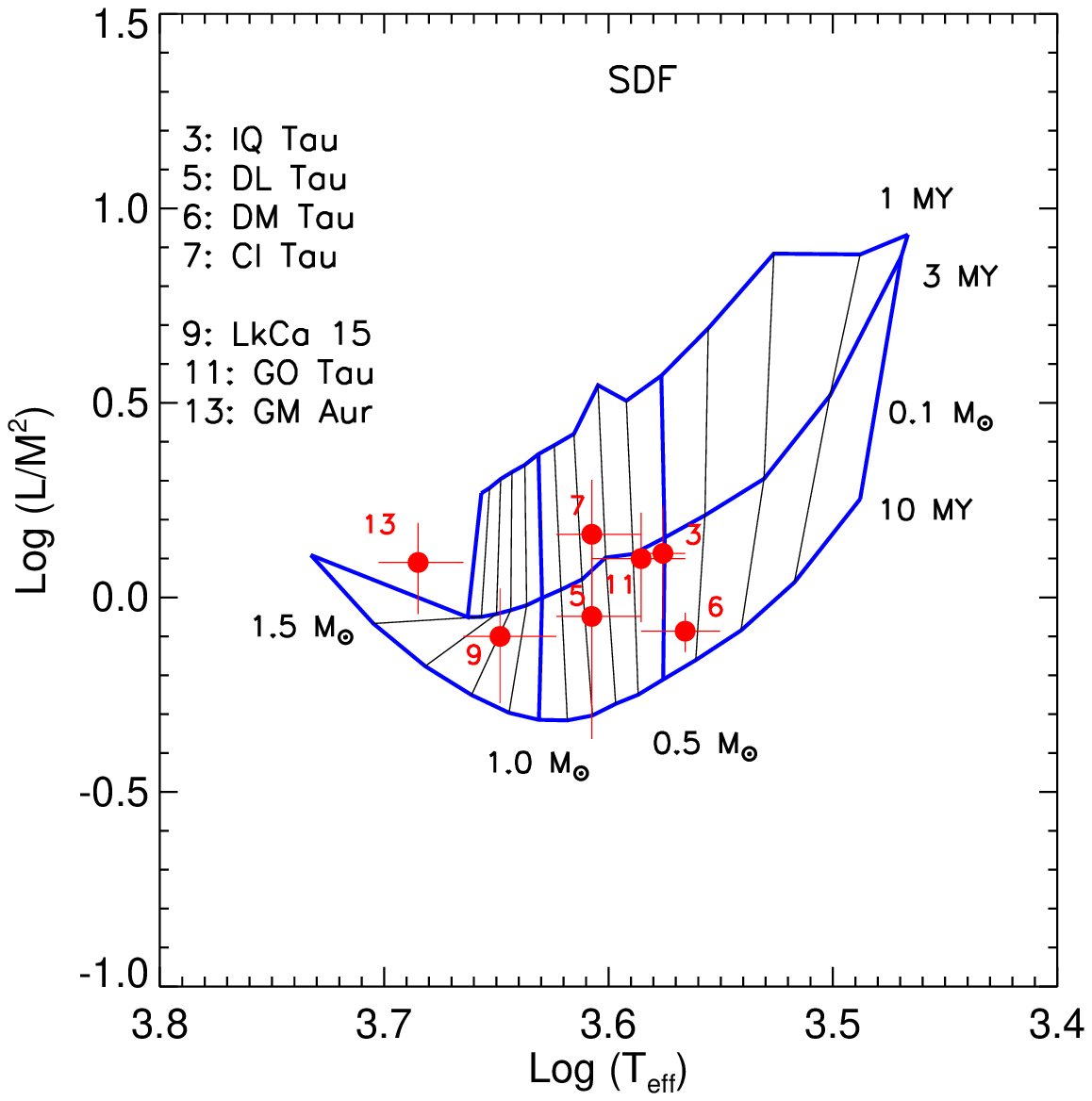}
\hspace{0.50cm}
\includegraphics[width=8.5cm]{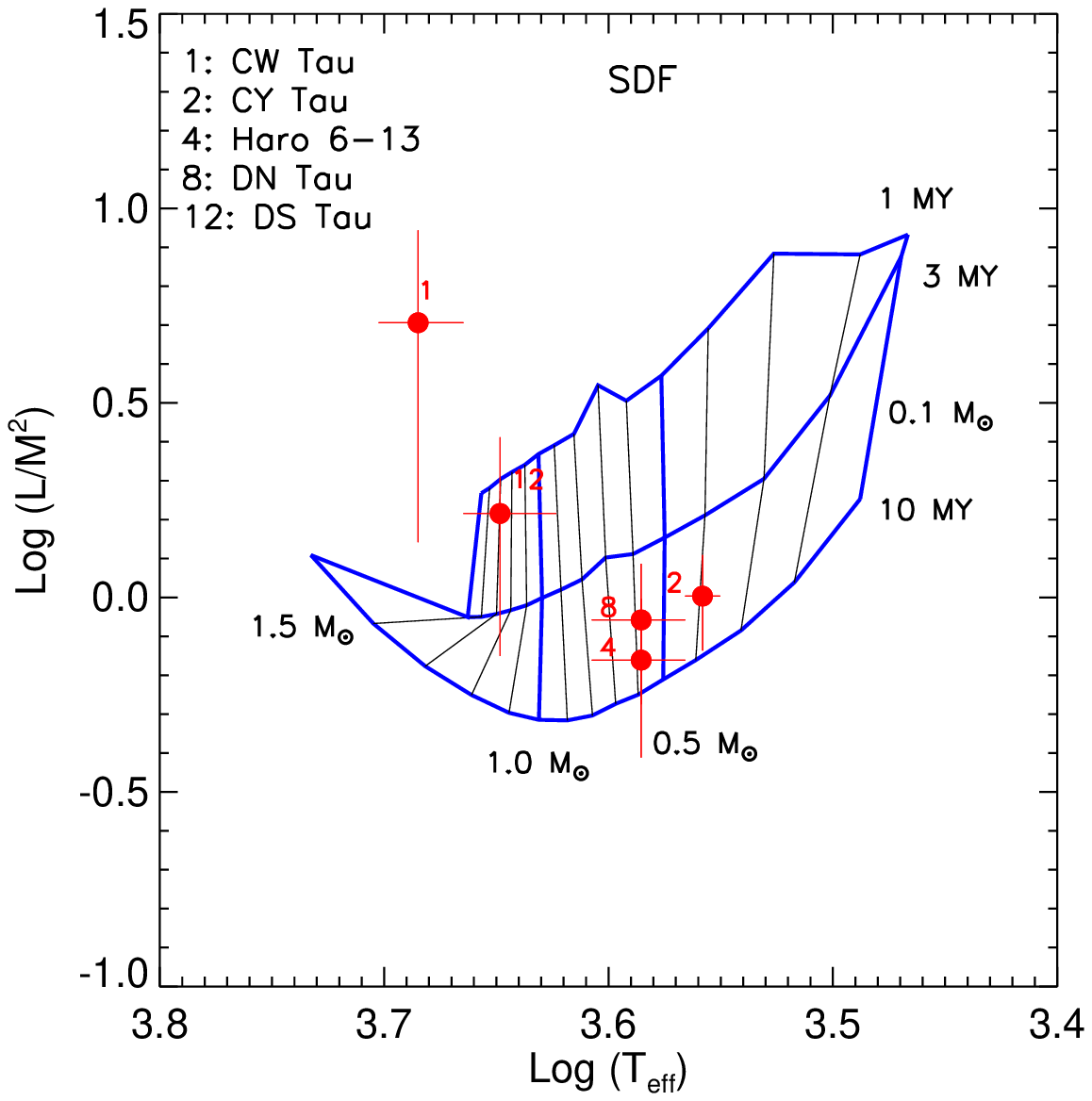}
\caption{As for Fig.\ref{fig:bcahtracks}, but 
for the \citet{Siess+etal_2000} tracks.}
\label{fig:siesstracks}
\end{figure*}

\begin{figure*}[h]
\includegraphics[width=8.5cm]{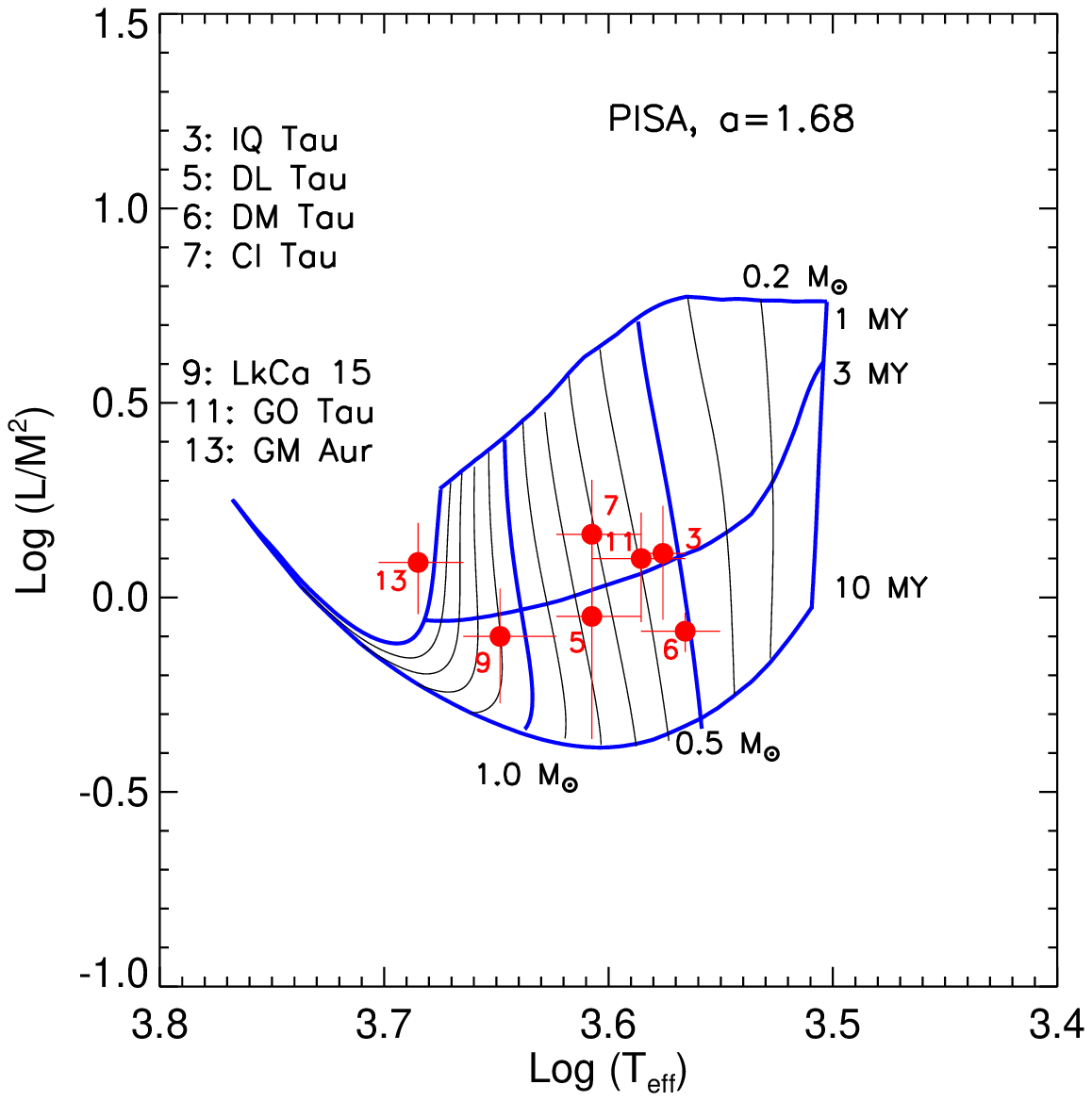}
\hspace{0.50cm}
\includegraphics[width=8.5cm]{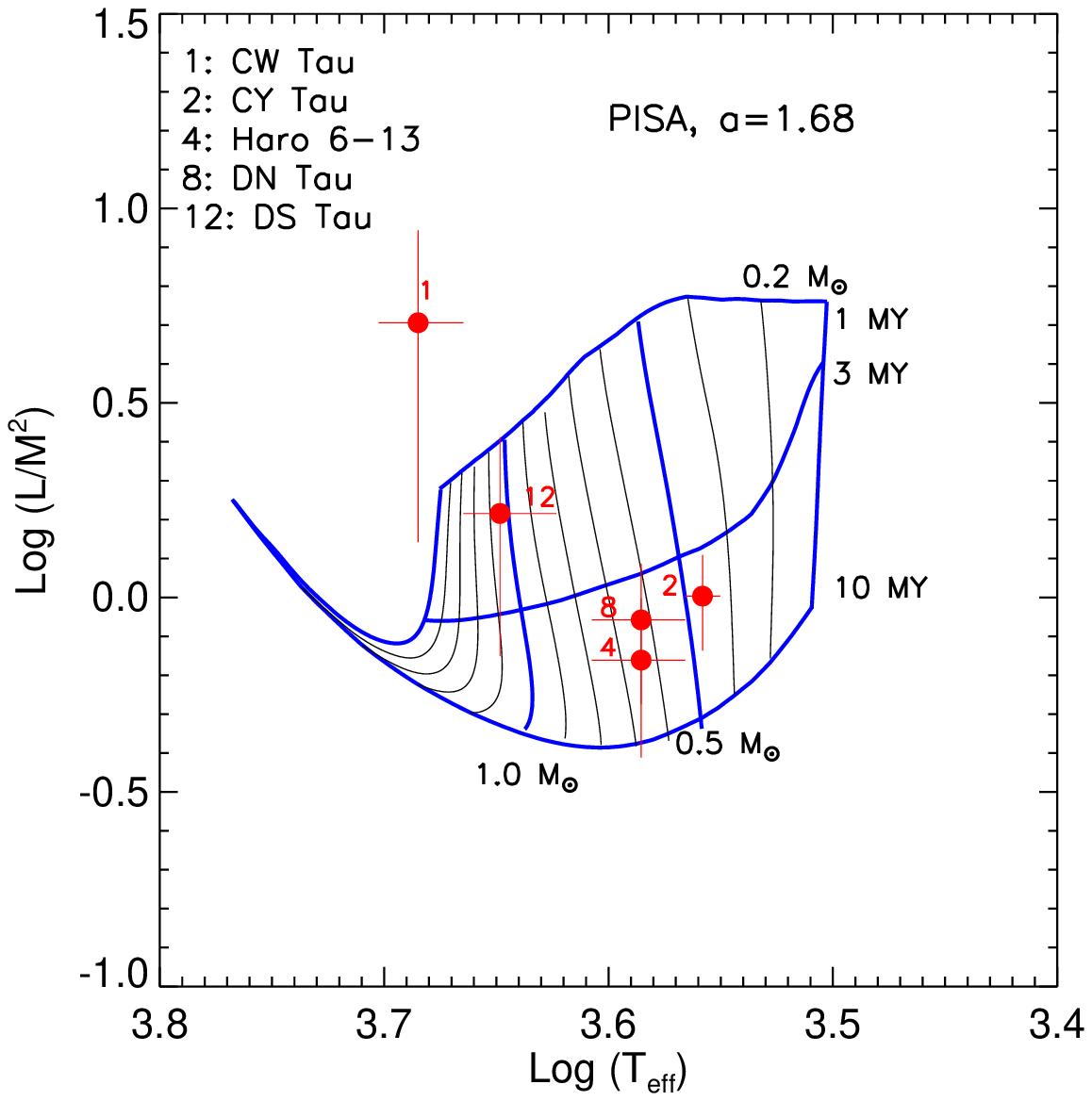}
\caption{As for Fig.\ref{fig:bcahtracks}, but
for the \citet{Tognelli+etal_2011} tracks.}
\label{fig:pisatracks}
\end{figure*}

To compare the masses we have measured to different evolutionary models,
we show in Fig.\ref{fig:bcahtracks}-\ref{fig:pisatracks} the position 
of the stars with accurate masses on  modified HR diagram 
for the \citet[][BCAH]{Baraffe+etal_1998}, the \citet{Siess+etal_2000} and the 
\citet[][PISA]{Tognelli+etal_2011} tracks; 
see also Fig.\ref{fig:dtmhtracks} for the Dartmouth \citep{Dotter+etal_2008} tracks. 
                   Taken together, the four different models differ
                   the most with respect to each other at masses less
                   than $0.5 \Msun$ and have different evolutionary time
                   scales for all masses at ages less than 5 Myr.


\end{document}